\documentclass[10pt,conference,letterpaper]{IEEEtran}
\makeatletter
\def\ps@headings{%
\def\@oddhead{\mbox{}\scriptsize\rightmark \hfil \thepage}%
\def\@evenhead{\scriptsize\thepage \hfil \leftmark\mbox{}}%
\def\@oddfoot{}%
\def\@evenfoot{}}
\makeatother
\usepackage{amsmath,amssymb,amsfonts,bbm,dsfont}
\usepackage[algoruled,medskip,dontprintsemicolon,linesnumbered,Algorithm]{algorithm}
\usepackage[noend]{algpseudocode}
\usepackage[anythingbreaks]{breakurl}
\usepackage{tikz}
\usepackage[hyphens]{url}
\usepackage{comment}
\usepackage{xspace}
\usepackage{multirow}
\usepackage{listings,color}
\usepackage{mathpazo}
\usepackage{soul}
\usepackage{multirow}
\usepackage{booktabs}
\usepackage{balance}

\usetikzlibrary{shapes,arrows,positioning}

\newfloat{algorithm}{thp}{lop}
\floatname{algorithm}{Alg.}

\newcommand{\Alg}[1]{Algorithm~\ref{alg:#1}}

\newcommand{\Hc}{H}
\newcommand{\Rc}{R}
\newcommand{\Ac}{A}

\newcommand{\pwdp}{P_{pa}}

\newcommand{\system}{\text{SoftATS}\xspace}

\definecolor{mygreen}{rgb}{0,0.5,0}

\makeatletter
\def\lst@makecaption{%
  \def\@captype{table}%
  \@makecaption
}
\makeatother

\begin{document}
	
\title{Software-Defined Adversarial\\Trajectory Sampling}

\author{}

\author{
 Kashyap Thimmaraju$^{1}$ \quad Liron Schiff$^2$ \quad \quad Stefan Schmid$^{1,3}$\\
{\small~$^1$ TU Berlin, Germany
\quad~$^2$ GuardiCore Labs, Israel 
\quad~$^3$ Aalborg University, Denmark}
}

\maketitle

\sloppy

\begin{abstract}
Today's routing protocols critically rely on the assumption 
that the underlying hardware is trusted.
Given the increasing number of attacks on
network devices, and recent reports on hardware 
backdoors this
assumption has become questionable. 
Indeed, with the critical role computer networks play today, 
the contrast between our security assumptions and reality is problematic.

This paper presents Software-Defined Adversarial Trajectory Sampling ($\system$),
an OpenFlow-based mechanism to efficiently 
monitor packet trajectories, also
in the presence of non-cooperating or even adversarial
switches or routers, e.g., containing hardware backdoors.
Our approach is based on a 
secure, redundant and adaptive sample distribution scheme which
allows us to provably detect
adversarial switches or routers trying to
reroute, mirror, drop, inject, or modify packets
(i.e., header and/or payload).
We evaluate the effectiveness of our approach in different adversarial settings,
report on a proof-of-concept implementation, and
provide a first evaluation of the performance
overheads of such a scheme.
\end{abstract}

\section{Introduction}\label{sec:intro}

Modern computer networks constitute
a crucial 
infrastructure: enterprise and
datacenter networks
as well as the Internet
in general need to provide high availability and
robustness. 
These increasingly stringent dependability requirements 
stand in stark contrast 
to today's vulnerable routing system.
In particular, while the problem of 
how
to provide authenticity and 
correctness of topology propagation and route computation
has been investigated intensively in the literature~\cite{secroute1,secroute2,secroute3,sbgp},
little is known about the vulnerabilities introduced
by an unreliable or even malicious/adversarial infrastructure~\cite{spook,chi2015detect,dhawan2015sphinx,kamisinski2015flowmon,eurosp17}.
Indeed, it seems challenging to perform routing over an unreliable infrastructure.

This is problematic:  
attackers have repeatedly demonstrated their ability to compromise
switches and routers~\cite{ciscoTelent,1001dsl,synful, lindner2009cisco,tpc-chairs},
thousands of compromised access and core routers are 
being traded underground~\cite{sats}, 
networking vendors have left backdoors open~\cite{huawei, netisbackdoor},
national security agencies
can bug
network equipment and introduce hardware backdoors~\cite{snowdencisco},
hacker tools scan and eventually exploit routers with weak passwords,
default settings
 are openly available on the Web, etc. 
The attack surface on network infrastructure is further 
exacerbated by vulnerable 
implementations of network protocols 
(e.g.~CVE-2014-9295, CVE-2014-0160, CVE-2015-6325),
especially in the context of more virtualized networks~\cite{kashyap-ovs},
as well as by today's trend toward
\emph{open networks}, rendering it
relatively easy for users to add switches and routers, establish 
links, and advertise routes. This concerns
both wireless and wireline networks (e.g., powerline,
overlay, or phone networks~\cite{sectrc}).

The problem is a fundamental one: even large enterprises
or national 
security agencies often cannot afford and do not have the expertise
to develop their
own trusted, high-performance network hardware.
Rather, they need to rely on equipment that
can be compromised during the supply chain~\cite{snowdencisco,stateOfItSec2015}.

An unreliable routing system introduces several threats:
for instance, wrongfully forwarded packets, i.e., packets following 
incorrect routes or ``trajectories'',  
may bypass security-critical components such as
firewalls or intrusion detection/prevention systems 
(e.g., in MPLS~\cite{icing}), and 
may be able
to enter or leave security critical zones. 
Forwarding or mirroring packets wrongfully
can also be used to violate isolation requirements in multi-tenant
datacenters (e.g., forwarding patient data between 
two health-care providers using
the same physical network~\cite{header-space}, which
is illegal~\cite{hipaa}), to infiltrate VPNs~\cite{sats}
or to exfiltrate sensitive information (see e.g.,
Operation Aurora~\cite{aurora}). 

While encryption may be used to mitigate
some of these problems, cryptographic approaches require 
an additional infrastructure, and also 
come with overheads at runtime. This is undesirable 
especially in high-performance networks.
Moreover, even encrypted traffic
may leak sensitive information, 
e.g., about the times and frequency of communications. 

Indeed, today, we lack good tools to verify routes 
in adversarial environments.  
For example, while 
traceroute and trajectory sampling tools 
are useful to verify routes in 
``reliable networks''~\cite{ts,bck-ts}, and
may still perform well in the context of faulty 
and heterogeneous networks~\cite{unrel-ts,csamp},
they are insufficient in non-cooperative 
environments: a compromised switch or router may
not reply with the correct information, 
and can also not conform to a scheme based on packet
labeling or tagging~\cite{pathquery,pathtracer}.

\subsection{Trajectory Sampling}
Trajectory Sampling (TS) is 
a low-overhead,
direct and passive measurement method
to infer packet routes.  
In a nutshell, in trajectory sampling,
packets are sampled
according to a random distribution at different
routers (usually defined by a hash over the immutable packet header space, 
e.g., the IP addresses and TCP ports), 
and sent to a \emph{collector}, 
reconstructing the trajectories based on the collected samples.
In general, in trajectory sampling, 
subsets of packets are sampled consistently: each
packet is sampled \emph{either at all routers it encounters, or at none}.
TS features a number of interesting properties:
\begin{enumerate}
\item It provides a direct observation of the
packets traversing the network. This is in
contrast to indirect
methods~\cite{feldmann2000netscope}
that involve uncertainties with respect
to the network's logical and physical state.

\item It is a passive measurement technique:
it does not require, for example probe packets
(which can introduce additional load and 
may be subject to a different
forwarding behavior),
nor does it require
the modification of packets 
(e.g., tagging requires
additional header space and rules at the routers).
It is hence a particularly attractive approach to monitor
the routes taken by packets for \emph{high-performance networks}
at a potentially low cost to forwarding.

\item It is a flexible technique that can be tuned for
the networking environment.
For example, the sampling rate can be adjusted
on demand and the load on the collector can
be adjusted accordingly.
Furthermore, the distribution used for sample
selection can be changed.
\end{enumerate}

However, a major shortcoming is that
classic TS relies on the assumption that all routers
are \emph{trusted}.
There is no mechanism preventing adversarial routers 
from misreporting sampling information.
As the sampling distribution is known to all the routers
(a packet is either collected at 
all routers it encounters, or at none),
an adversarial router could simply exploit knowledge about
gaps in the sampling space to conceal misbehavior.
For example, an adversarial router can drop all
packets not being sampled and is guaranteed to go
undetected unless the sampling distribution is changed.
Hence what we need is
a solution which tolerates adversarial routers,
prevents exploiting holes in the sampling space, and
detects forwarding based attacks such as packet
drops and injections.

To prevent routers from exploiting holes in
the sampling space, Lee et al.~\cite{sats}
propose routers to sample different packets with
pairs of routers sampling the same packet.
They call their scheme ''secure split assignment''
trajectory sampling. Although novel, and
dare we say ahead of their time, the authors
only present a first simulation of detecting
packet drop related attacks.
As we will shortly see, our
scheme \emph{Adversarial Trajectory Sampling (ATS)}
develops their idea further.

\subsection{Contributions}

This paper makes the following contributions:
\begin{itemize}
\item We identify and model a wide range of routing
attacks that can be launched by untrusted data plane
components such as \emph{drop}, \emph{injection},
\emph{denial-of-service}, \emph{man-in-the-middle},
\emph{rerouting}, \emph{reconnaissance}, \emph{mirroring}
and \emph{modification} attacks.

\item We observe that 
Software-Defined Networks (SDNs) provide an ideal framework to perform 
secure trajectory sampling in adversarial environments.

\item We present parallelizable algorithms that leverage \emph{secure
and redundant} dynamic sampling schemes to detect routing
attacks from a logically centralized controller.

\item We formally prove the effectiveness of our detection
algorithm by deriving the detection probabilities for
different attacks.

\item We present and evaluate an OpenFlow based prototype.
\end{itemize}

\subsection{Paper Scope}

In general, the goal of our approach is
to empower the network operator to \emph{detect} misbehavior, 
as opposed
to \emph{prevent} misbehavior. In other words, alone, our approach
is unable
to ensure a packet will not traverse certain
network regions or reach certain destinations.  However, we 
believe that the possibility
to detect misbehavior is a strong incentive for 
routers and vendors to not deviate from the correct behavior.
Moreover, we in this paper do not consider the orthogonal
question
of how a user should specify its desired and undesired
routes to the network operator. 

\subsection{Organization}

The remainder of this paper is organized as follows.
Section~\ref{sec:model} introduces our threat model.
Section~\ref{sec:system} and 
Section~\ref{sec:softatssys}
present the challenges and our proposed solution in detail.
Section~\ref{sec:analysis} 
analyzes the detection probabilities achieved by our approach.
Section~\ref{sec:prototype} reports on our prototype implementation,
and Section~\ref{sec:eval} presents experimental results on the detection
time and performance overheads under
real traffic workloads.
Section~\ref{sec:discussion} discusses additional aspects and extensions
of~$\system$.
After reviewing related literature in Section~\ref{sec:relwork},
we conclude our work in 
Section~\ref{sec:conclusion}.

\section{Threat Model}\label{sec:model}

We consider a network consisting 
of a set of \emph{switches} (or for the purpose
of this paper equivalently: routers), 
connected by a set of 
\emph{links}. 
We consider an adversarial model
where switches
are untrusted,
e.g., they may contain (hardware and/or software)
backdoors that may be introduced
by compromising the vendor's supply chain.
Accordingly, we do not place any restrictions on what an
adversarial switch can and cannot do.
For example, an adversarial switch can fabricate and transmit 
any type of message, both in the data plane (e.g.,
duplicate packets)
as well as in the control plane (e.g., report wrong
samples); it can also arbitrarily deviate 
from the OpenFlow specification,
all at the risk of being detected.
However, we assume that the network edge (e.g., 
the servers in the datacenter) are trusted.
This assumption is necessary for the following reason:
If a packet only traverses the edge, then
it is \textbf{impossible}
to detect whether the edge switch forwarded
the packet correctly or not using our scheme.

\begin{figure*}[t]
	\includegraphics[width=.45\textwidth]{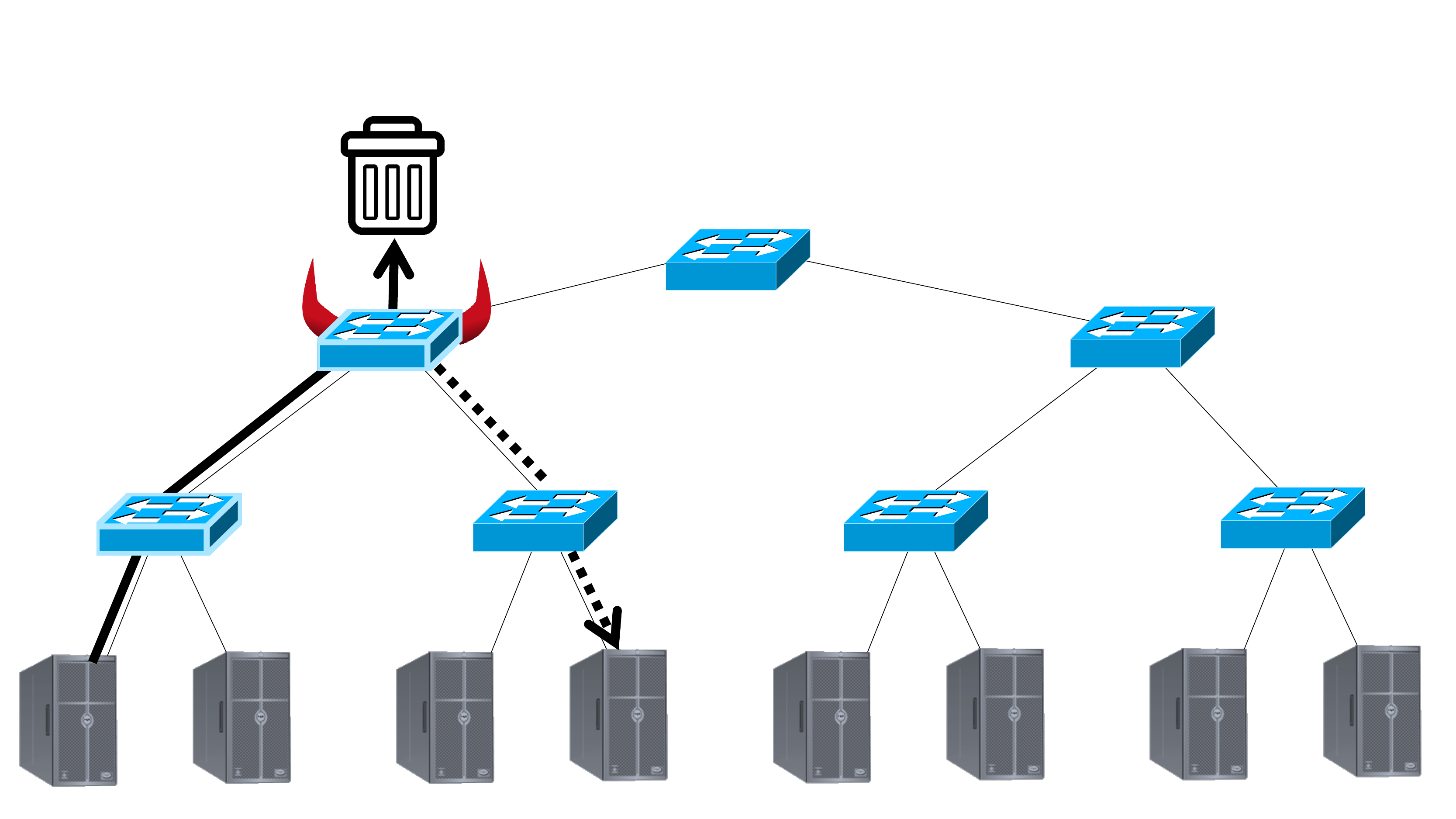}
	\includegraphics[width=.45\textwidth]{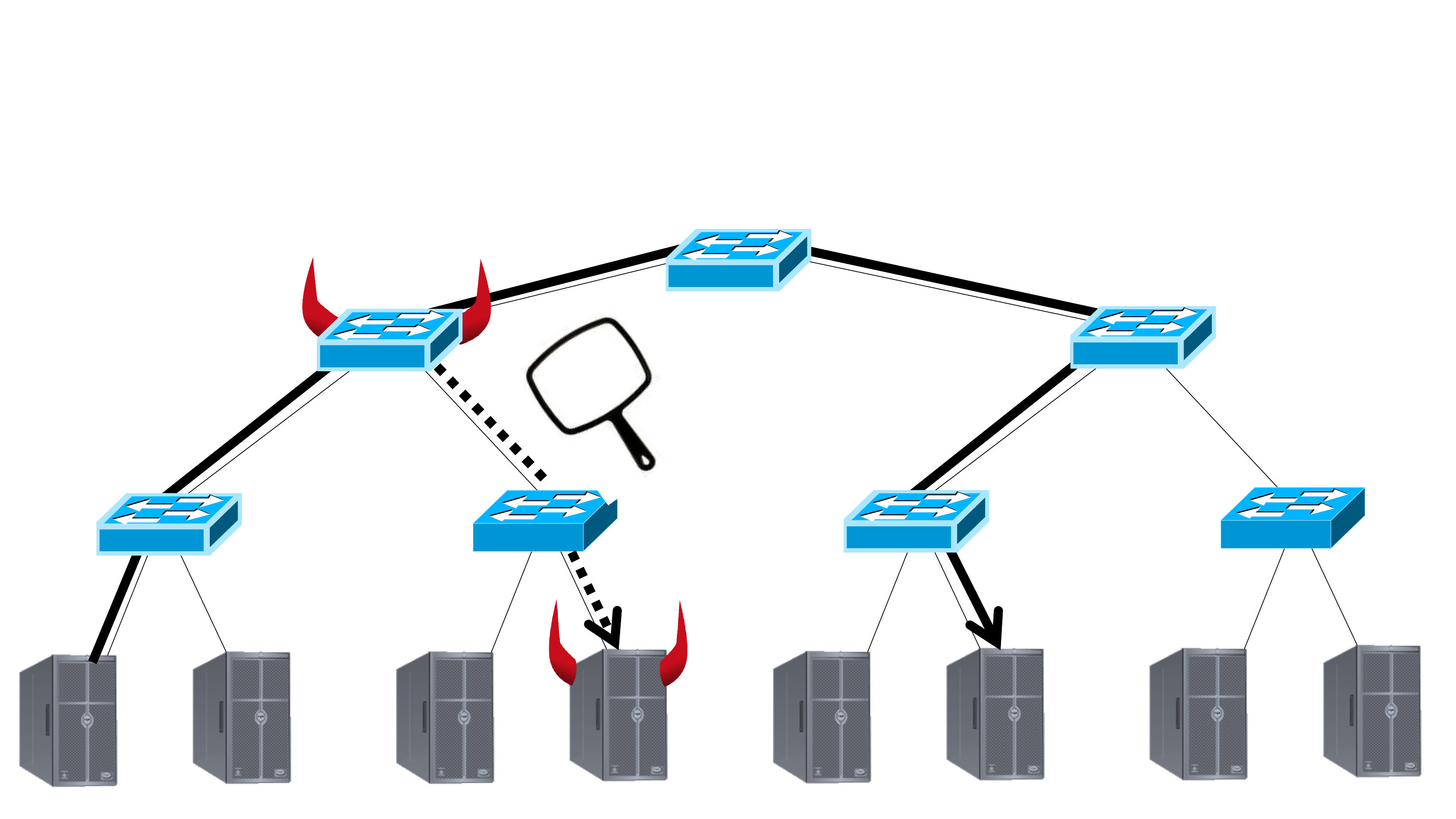}\\
\includegraphics[width=.45\textwidth]{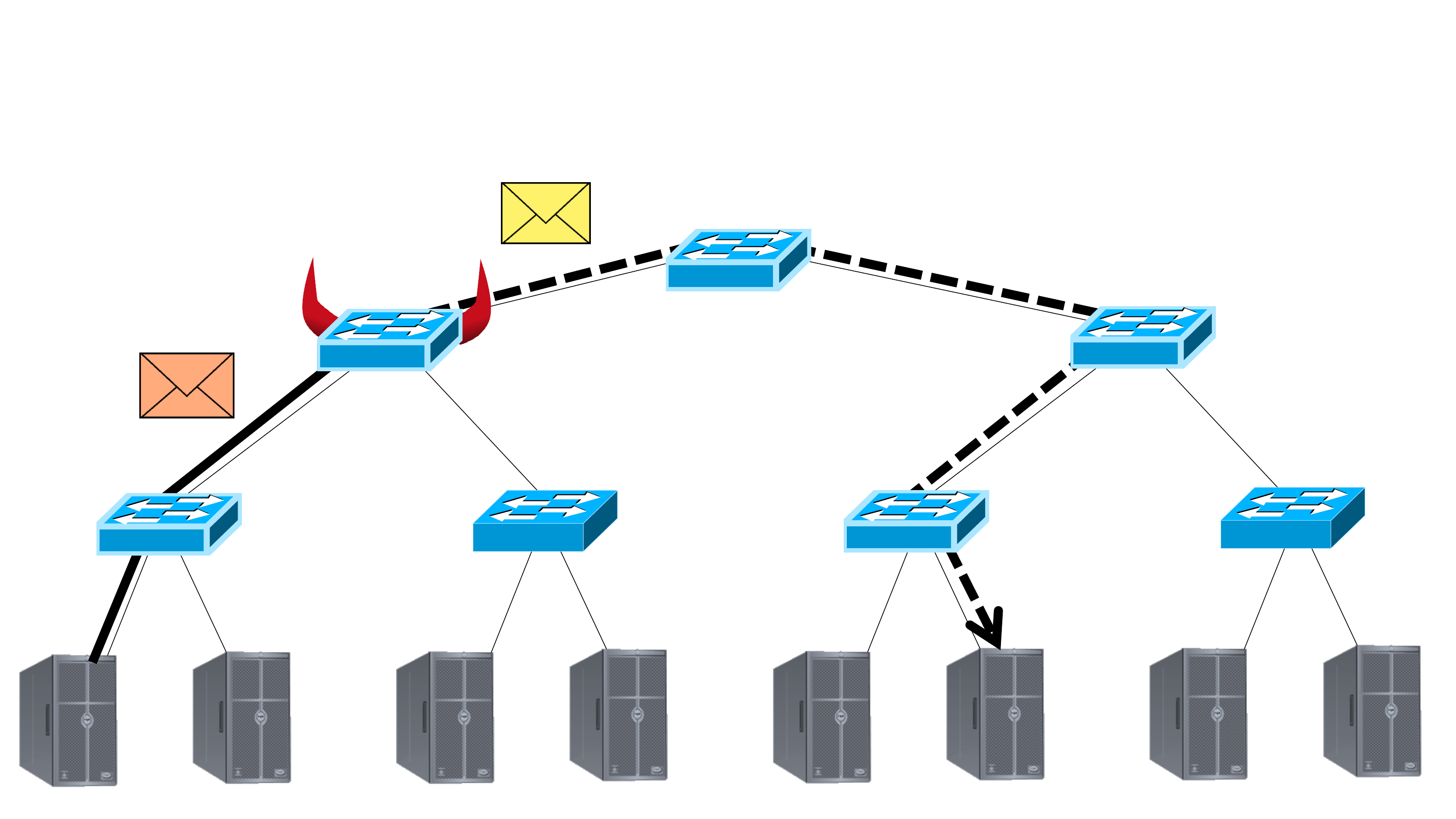}
\includegraphics[width=.45\textwidth]{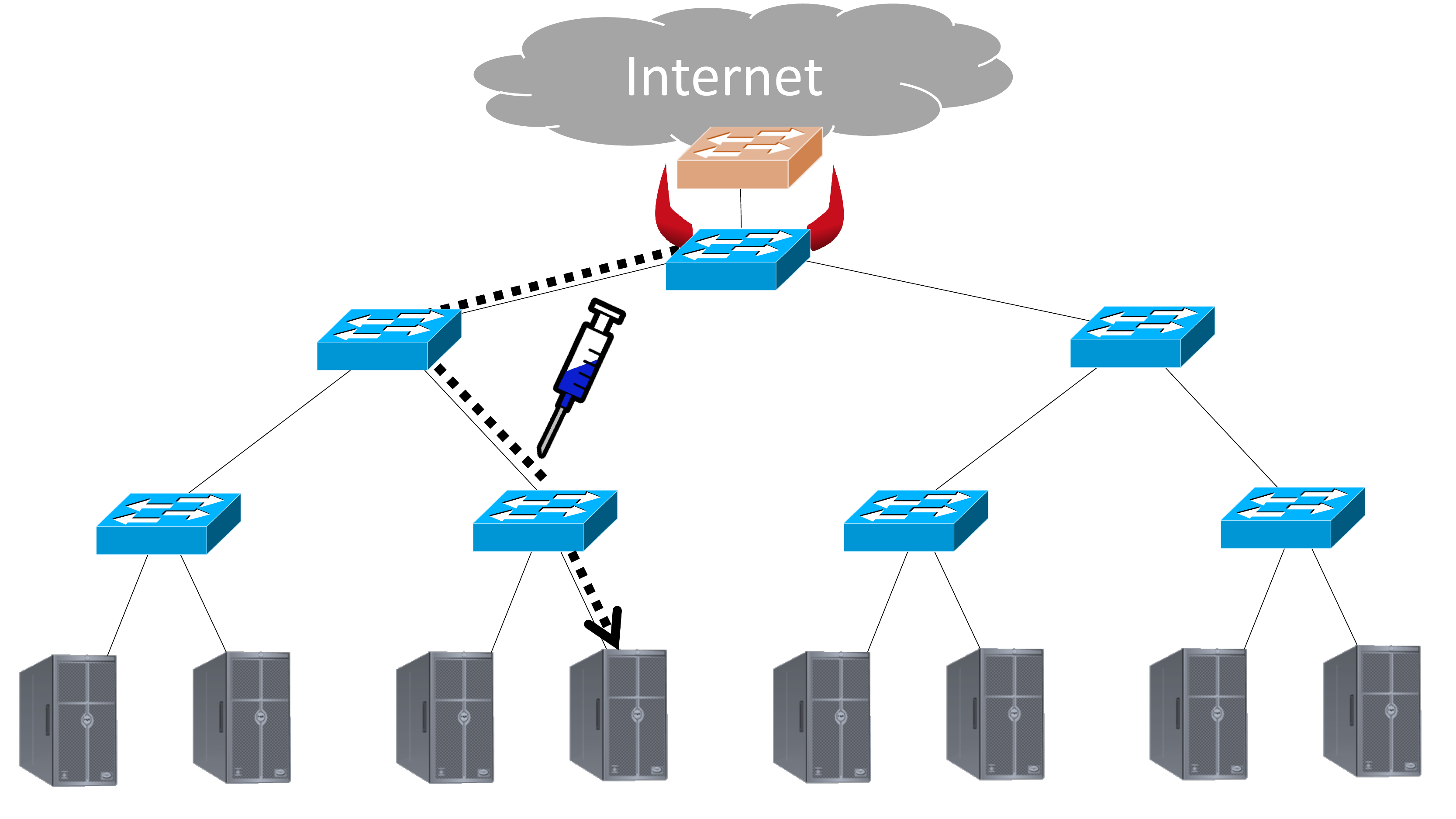}
	\caption{Overview of possible malicious switch attacks. 
As an example, a Clos (``fat-tree'') topology is depicted
(for ease of representation, we aggregate links in this figure, 
see~\cite{clos} for a full representation): servers are organized
into racks, and are interconnected via so-called \emph{Top-of-Rack (ToR)}
switches. Racks are connected by aggregation switches to
form \emph{pods}.  Finally, pods are connected by core switches,
which may also connect the datacenter to the Internet.
  \emph{Top left:} Denial-of-service attack resp.~packet drop: instead of forwarding
	the packet to the server in the second rack (\emph{dashed path}), the malicious switch drops the packet.  \emph{Top right:} The malicious switch injects a copy of the packet to 
the rack (\emph{dashed path}), in addition to sending it along the regular path (\emph{solid path}). 
In the rack where the packet is mirrored to, a malicious server may exfiltrate confidential information.
 \emph{Bottom left:}  A malicious switch modifies the 
 packet along the route (man-in-the-middle attack). 
 \emph{Bottom right:} A malicious core switch
 injects a harmful packet to attack an internal server (an insider attack).}
	\label{fig:attacks}
\end{figure*}

More systematically, an adversarial switch may
perform the following attacks (cf.~Figure~\ref{fig:attacks}).  
\begin{enumerate}
\item \textbf{Denial-of-service:}
It can drop transit packets.
\item \textbf{Rerouting:}
It can forward a packet to the
wrong port (e.g., breaking logical isolations).

\item \textbf{Mirroring:} 
It can duplicate a packet,
and e.g., send one to the correct
and one to an incorrect port.

\item \textbf{Man-in-the-middle:} 
It can also delete packets,
generate new packets, or modify the header
or payload of packets (e.g., changing
the VLAN to break isolation domains).

\end{enumerate}

In fact, we observe that all attacks considered in this
paper can be represented by the following two primitives:

\begin{enumerate}
\item \textbf{Drop:} An
adversary chooses a specific source-destination pair,
and drops some (or all) packets communicated
between this pair. 
\item \textbf{Injection:}
An adversary injects new packets (or packets sent
earlier to other destinations) to the network. 
\end{enumerate}

\begin{table}[h]
\centering
\caption{A comprehensive overview of attacks, their respective primitives, and a short description of the attack.}
\label{tab:attackTypes}
\footnotesize
	\begin{tabular}{p{2cm} p{2cm} p{3.2cm}}
	\toprule
	\textbf{Attack}&\textbf{Attack primitive}&\textbf{Attack description} \\ \hline
	\midrule
	Denial-of-service&Drop&Switch simply drops the packet(s).\\
	Reconnaissance&Injection&Switch simply injects a new packet.\\
	Man-in-the-middle&Drop+Injection&Switch modifies the original packet \textbf{and} forwards it along the original path.\\
	Man-in-the-middle proxy&Drop+Injection&Switch drops the original packet \textbf{and} forwards a modified version along a new path.\\
	Mirror	&Injection&Switch forwards the original packet \textbf{and} sends a modified version along a new path.\\
	\bottomrule
	\end{tabular}
\end{table}

For instance, packet re-routing attacks,
where an adversary reroutes a packet from some source to
an illegitimate destination, can be modelled as 
a combination of two independent
attacks where in the first attack the packet is dropped
and in the second the packet is injected. 
Mirroring can be described as an 
injection of copied packets at some switch. 
A modification can be modelled as
dropping and subsequently injecting
modified packets (e.g., different header or payload). 
Packet delays can be abstracted as a drop and an injection attack.

Note that drops and injection are not only primitives
resp.~building blocks of more complex attacks, 
but can also be used as standalone attacks: 
frequent drops result in a denial-of-service, 
and injection could be used to contact 
Command-and-Control (C\&C) servers of a botnet.
Table~\ref{tab:attackTypes} presents the attacks,
attack primitives, and the attacks description.

There may be more than one adversarial switch,
and adversarial switches may even collude:
e.g., for covert switch-to-switch communication, one switch 
can inject a packet and the other drop it; or, one
switch does not report the to-be-sampled packets of the colluding switch,
or reports a non-existent packet.
However, obviously and as we will see, 
the detection probability of our approach
decreases with an increasing number of adversaries. 
Accordingly, we in the following, assume that the number
of malicious and colluding switches is small. 
We argue that the assumption
that only a subset of switches are malicious and collude, is reasonable
in practice: for example, switches from different vendors
or switches manufactured in different countries,
are unlikely to collude.
Indeed, leveraging heterogeneity is a key principle to
improve security of networks~\cite{achenbach2014universally,disn16netco}.

The solution proposed in this paper relies
on a software controller (namely the OpenFlow controller)
which securely distributes sampling rules and collects samples.
Accordingly, we assume that this controller and its applications
are trusted entities:
for example, they are
developed in-house and
audited using
static and
dynamic 
program analyses---an affordable measure for
large enterprises or security agencies.
Moreover, we show that our algorithms are 
parallelizable, enabling a
highly available implementation of these components.

\section{The Case for an SDN Approach}\label{sec:system}

We first identify
the fundamental challenges and requirements
of \emph{Adversarial Trajectory Sampling (ATS)}, and then
describe how opportunities in SDN meet the stated requirements.

\subsection{Requirements and Challenges}

In general, we can decompose the 
challenges involved in designing a strong adversarial
trajectory sampling scheme into several
sub-challenges:

\noindent \textbf{Computing and Assigning Redundant Sample Ranges.} 
The sampling strategy of $\system$ is based on the split assignment approach 
by Lee et al.~\cite{sats}, which requires the computation
and secure distribution of ``good sample ranges''.
These ranges should be allocated
\begin{itemize}
\item \emph{$\ldots$ independently at random:} given its own sampling ranges,
a switch cannot guess with high probability the location or
the sampling range of other switches.
\item \emph{$\ldots$ sampling space:} not all switches should sample
the same packet. Instead, different switches should sample
different packets.
\item \emph{$\ldots$ redundantly:} although different switches
should sample different packets, there must be an 
overlap so that misbehaving switches can be
detected by other switches.
\end{itemize}

There are different approaches to achieve
such a sampling distribution. For example,
for each hash range,
a fixed number (e.g., two) of switches may be selected
uniformly at random in the network.
Alternatively, samples may be assigned in a more flexible
manner, without imposing strict constraints on the
number of replicas.

\noindent \textbf{Securely Distributing Samples.} 
A mechanism is needed to securely distribute samples
to the switches. The samples must be encrypted to
preserve confidentiality, and authenticated
to preserve data integrity.
However, this may be challenging when
the bi-directional
channel between the sample
generator and switch are not connected via
a dedicated network (commonly called
the out-of-band network or management network).
Malicious switches can drop packets to the
controller that contain the samples.

\noindent \textbf{Parallel Analysis.}
Interpolating trajectories as well as detecting anomalies
requires computational resources on the collector.
In order to sustain high performance workloads,
the detection logic should be parallelized and distributed. 

\noindent \textbf{Avoiding Biases.} 
The sampling should affect all packets with equal probability.
Moreover,
only an unbiased sampling can guarantee
a maximal detection probability: if certain packets (e.g.,
packets with certain source or destination addresses)
are sampled more frequently than others, this
may be exploited by an attacker.

\noindent \textbf{Dynamic Reassignments.} 
Trajectory sampling, with or without split assignments,
only covers a subset of packets. In order to maximize the detection probability
and prevent adversaries from learning monitoring patterns,
and while keeping the solution scalable, sampling distributions
should be changed over time. However, this is non-trivial, as
changing the assignments on the switch while traffic
is sampled yields synchronization constraints~\cite{7542179,Afek:2014:RCC:2620728.2620780}.

\subsection{Making The Case for SDN}
            
We find that Software-Defined Networks (SDNs) in general
and OpenFlow in particular
provide an ideal
environment to implement an adversarial
trajectory sampling for the following reasons:

\begin{enumerate}
\item \emph{Programmable, logically centralized control:}
The OpenFlow controller provides an ideal platform
to implement the logic of the sampling and the collector: 
simple controller applications can be used to compute smart sample 
range distribution schemes, and to distribute, collect and analyze trajectories.

\item \emph{Secure communication channels:} An encrypted
and authenticated channel
between switches (i.e., OpenFlow switches) and controller is 
readily supported in OpenFlow. These connections are based
on TLS/SSL, include sequence numbers and could also be established in-band.

\item \emph{Support for sampling:} An OpenFlow switch readily supports
the necessary functionality for sampling. That is,
a sampling range can easily be added, changed, and removed at the switch,
in the form of \emph{flow rules}. Moreover, OpenFlow switches such as Open vSwitch
readily support computing (simple) hash functions~\cite{ovs}. 

\item \emph{Support for taking into account payload:} Unlike legacy switches,
OpenFlow switches can operate not only on Layer-2, but also Layer-3 and
Layer-4 header fields. Accordingly, sampling can be performed
on a per-packet granularity rather than a per-flow granularity. 
For example, sample decisions may be based on the TCP checksum
field which depends on the packet payload. Moreover, some OpenFlow
switches (such as NoviKit$^{\text{TM}}$) 
also support sampling based on packet payloads (e.g., 
defined by a certain offset).
As we will see, by collecting
entire packets and not just headers (as readily
supported by the \emph{Packet-in} mechanism), 
more sophisticated attacks can 
be prevented.
\end{enumerate}

\begin{figure}[t]
	\centering
	\includegraphics[width=1.0\columnwidth]{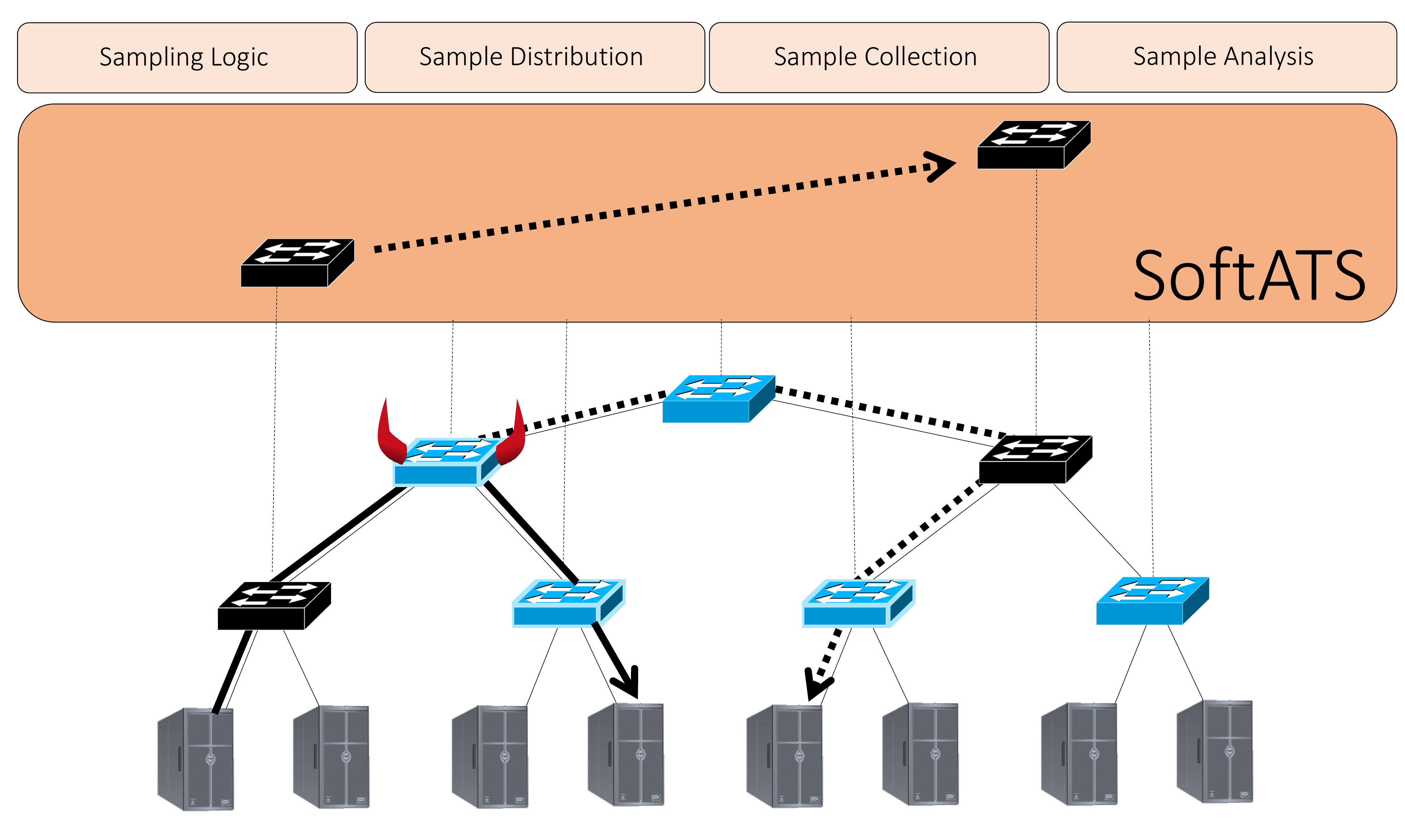}
	\caption{Overview of~$\system$: Switches
	are connected to a logically centralized controller,
	which collects samples (\emph{here:} from the two \emph{black} switches)
	in order to observe trajectories. In this case, the samples provided
	by the black switches reveal that a malicious switch must have mirrored
	the packet (intended route indicated as \emph{solid line}, mirrored extra path indicated as \emph{dashed line}).}
	\label{fig:overview}
\end{figure}

\section{The SoftATS System}\label{sec:softatssys}

In this section, we describe our Adversarial Trajectory Sampling (ATS) scheme
system which provably detects a wide range
of routing attacks in SDNs. 
In particular, we
present the $\system$ architecture and our algorithms in detail.

$\system$ is designed for an OpenFlow-based network
and a logically centralized (but possibly physically distributed)
controller. 
The~$\system$ controller serves as
the (secure) distributor of sampling ranges,
as the (secure) collector
of the sampled packets,
as well as the analyzer accordingly. 
As we will discuss later in more detail, some of this logic
can easily be parallelized.~$\system$ relies on
the following concepts: 
\begin{itemize}
\item \textbf{Packet Hashing Function~$h$:}  A hash function 
$h:\{0,1\}^*\rightarrow \Hc$ which maps (parts of) the packet
to some \emph{hash domain}~$\Hc$. 
We consider two types of hashing approaches:
\begin{enumerate}
\item \emph{Header based:} 
The hash is computed only over the immutable header fields of the packet.
\item \emph{Payload dependent:} The hash also depends on the
 payload of the packet.
\end{enumerate}
\item \textbf{Hash Assignments~$\Ac(s)$:} Each 
switch~$s$, is assigned 
a set of hash values~$\Ac(s)$. The switch is configured to report (sample) 
any packet with content that hashes to~$\Ac(s)$. 
The size of~$\Ac(s)$ divided by the size of the hash domain 
defines the \textbf{sampling ratio} of switch~$s$, 
denoted by~$p_s$. In other words:~$$p_s = \frac{|\Ac(s)|}{|\Hc|}$$

\item \textbf{Assignment Granularity~$\delta$:}
The set~$\Ac(s)$ can be configured in the switch using 
a set of configuration 
rules~$\Rc(s)$, each corresponding to a disjoint subset of~$\delta$ hash values, 
e.g., through the use of a \emph{range} or a 
\emph{ternary bit pattern}. 
The following equality holds:~$$|\Rc(s)|\cdot \delta = |\Ac(s)|$$ 
Clearly the granularity affects the configuration space complexity.

\item \textbf{Redundancy:}
In order to verify the trajectory of a sampled packet in the network, 
it must be sampled at more than one switch (but for secrecy purposes, not all). 
\end{itemize}

\system is a modular framework 
which supports many different sample assignment 
strategies, leveraging SDN's flexible configuration and packet matching capabilities. 
Next we discuss 
four important elements of the adversarial 
sampling process
in turn:~$(i)$ how hash assignments are applied in the switches,
$(ii)$ how hash assignments can be changed over time,
$(iii)$ how sampled packets are collected, and
$(iv)$ how to reconstruct the trajectories given the samples, 
using a highly parallelizable algorithm.

\noindent \textbf{Configuring Hash Assignments for Sampling.}
\system offers many flexibilities in terms of assignment. 
For example regarding the hash assignment for each switch in the network,
there are different approaches, e.g.: 
\begin{enumerate}
\item \emph{pairwise:} 
For each pair of switches,~$s_i,s_j\in S$, we assign 
a randomly selected subset of hash values~$\Ac(s_i, s_j)$. 
In total, each switch~$s_i$ is assigned with the union of all subsets 
selected for all its pairs, i.e.,~$\Ac(s_i)= \cup_{j\neq i}\Ac(s_i, s_j)$.
\item \emph{independent:} Each switch is assigned with a randomly 
selected subset of hash values, independently from other switches.
\end{enumerate}

In the pairwise approach,
we also ensure multiple matches (with probability 1), 
avoiding ineffective assignments (e.g., with local unmonitored ranges) 
that might be chosen in the independent assignment.

The hash function can be applied to just the packet header or 
the payload as well. In general, \system maintains an OpenFlow session to each switch 
and uses it to send \emph{Flow-mod} or 
\emph{Group-mod} commands according to the switch specific hash assignment.

\noindent \textbf{Packet Header Hashing.}
In case the hash function should be applied to the packet header,
we configure the default hash based selection group 
with~$|\Hc|/\delta$ buckets. Among these buckets,
$|\Rc(s)|=|\Ac(s)|/\delta$ buckets are defined to send the packet to the collector.
The indices of the sampling buckets are chosen according 
to the assigned hash values. Note that by using weighted buckets, 
the total number of buckets can be linear in~$|\Rc(s)|$, e.g., 
replacing~$k$ consecutive dummy buckets with one dummy bucket of weight~$k$.
An example of such a configuration is shown in Listing~\ref{header}
\begin{lstlisting}[float=*,frame=single, label=header, caption=An example OpenFlow group and flow table used to sample packets using the header hashing. A vlan tag is pushed into the packet to determine the bucket the packet was hashed to.]
group_id=1,type=select,bucket=actions=push_vlan:0x8100,set_field:1234->vlan_vid,CONTROLLER:65535,
bucket=actions=push_vlan:0x8100,set_field:1235->vlan_vid,CONTROLLER:65535
cookie=0x1, table=0, priority=1,in_port=2,nw_dst=1.1.1.0/24 actions=output:3,group:1
\end{lstlisting}

\noindent \textbf{Packet Payload Hashing.}
In case the hash function should be applied
to the packet payload,
then the switch is configured to match the 
TCP/UDP checksum field. 
Matching the checksum field using OpenFlow,
requires the experimenter extension 
and has already been prototyped by Afek et al.~\cite{liron-sampling}. 
We emphasize that TCP/UDP checksums are only used at the switch 
for sampling, while deeper payload verification (hashing) can be performed at
the collector.
Matching the TCP/UDP checksum field alleviates the
overhead of checksum computation on the switch.
The flexibility of OpenFlow allows custom
match fields, therefore for packets that do not
include checksum fields, a custom match field and
hash algorithm may be used.
Concretely, assuming that the rules (subsets)~$\Rc(s)$ can be expressed by 
ternary patterns,~$s$ is configured with 
exactly~$|\Rc(s)|$ flow entries, each sending 
the matched packet to the collector.

In both cases, 
sending the sampled packets to the collector is performed by 
the OpenFlow action to forward to the controller as a
\emph{Packet-in}.
An example of such a configuration is shown in Listing~\ref{payload}
\begin{lstlisting}[float=*,frame=single, label=payload, caption=An example OpenFlow flow table used to sample packets using the tcp checksum field.]
cookie=0x1, table=0, priority=2,tcp,chksum=1234 actions=CONTROLLER:65535,goto_table:1
cookie=0x2, table=0, priority=1,ip actions=goto_table:1
cookie=0x3, table=1, priority=1,in_port=2,nw_dst=1.1.1.0/24 actions=output:3
\end{lstlisting}

\begin{algorithm}[t]
	\begin{algorithmic}[1]
		\Require sampling rate~$p$, pattern size~$\delta$, total hashes~$\Hc$, switches~$S$
		\State~$x\gets p\cdot|\Hc|/\delta\cdot(|S|-1)$ \Comment\emph{number of hash patterns per pair of switches}
		\State~$assignedHashes\gets 0$ \Comment\emph{initialize the assigned hash counter}
		\State~$totalHashes\gets x\cdot|S|\cdot(|S|-1)$ \Comment\emph{number of hash patterns for all switches}
		\While {$assignedHashes < totalHashes$}
			\State {\bf randomly choose} a switch-pair $(s_i,s_j)_{i \ne j}\in S$
			\State {\bf randomly choose} a hash pattern~$h\in\Hc$
			\If {$|\Ac(s_i, s_j)| < x \&\& |\Ac(s_j, s_i)| < x$}
            	\State~$\Ac(s_i, s_j).add(h)$
            	\State~$\Ac(s_j, s_i).add(h)$
				\State~$assignedHashes \gets assignedHashes+2$
			\EndIf
		\EndWhile
		\Return hash assignments~$\Ac$
	\end{algorithmic}
\caption{Initial Pairwise Assignment of Hash Values
	\label{alg:assignment}}
\end{algorithm}

\noindent \textbf{Dynamic Configuration.}
It is useful to change the hash assignment on the
switches as
at any moment in time,
$\system$ samples only a fraction ($p_s$) of the traffic.
This way, over time,
it becomes difficult for the adversary to avoid
being detected as no information about a static
sampling pattern is leaked.

However, changing the hash assignment over time is non-trivial.
During the addition or removal of a hash assignment from 
a switch configuration, there is an uncertainty regarding 
the exact time at which an update takes effect:
the communication network between the controller
and the switch is inherently asynchronous, and also
data structure reconfiguration times at the switches
may differ~\cite{kuzniar2015you}.
To avoid false positives, 
\system needs to temporarily 
suppress alerts related to the updated 
switch and hash values.
In particular, hash values should be changed at \emph{random times},
and involving only one (pair of) switch(es) at a time: thus, 
a malicious switch cannot abuse the time period
of uncertainty.
A simple update scheme is described in \Alg{permutation}.

\begin{algorithm}[t]
	\begin{algorithmic}[1]
		\Require switches~$S$, total hashes~$\Hc$, hash assignments~$\Ac$
		\While{True}
		\State {\bf randomly choose} a switch pair~$(s_i,s_j)_{i\neq j}\in S$
		\State {\bf randomly choose} a hash index~$k\in\{1..x\}$
				\State {\bf randomly choose} a hash~$h\in\Hc$
		\State~$\Ac(s_i,s_j)[k]\gets h$
		\State {\bf wait} a random period
		\EndWhile
	\end{algorithmic}
	\caption{Random Permutation of Hash Assignment Values
		\label{alg:permutation}}
\end{algorithm}

\noindent \textbf{Collecting the Samples.}
Reporting the samples from all switches to the collector is 
performed using \emph{Packet-in}s within the encrypted 
and authenticated OpenFlow session, 
protecting them even if communication is 
in-band; this addresses known sample integrity issues~\cite{sats}. 
Nevertheless, whenever possible, we suggest using out-of-band control 
in order to prevent sophisticated attacks, e.g., 
using burst analysis of the inband control channel.
In case of distributed control planes, 
the \emph{Packet-in}s (the samples) can be sent to the same 
controller which configured the sampling rules, or to another controller server 
depending on operator requirements. 

\noindent \textbf{Analyzing the Samples.}
At the heart of trajectory sampling lies the 
construction or ``interpolation'' of trajectories out of samples.
In an SDN, the controller has a global view of the network,
i.e., it knows the location of hosts in the network,
the switches, links between switches,
and links between switches and hosts. 
For example, with two samples from two different switches,
the controller can compute and compare the packet's
trajectory using the topological information it has.
This makes it ideal for trajectory computation.
Based on the collected samples, 
anomalies such as missing or extra samples are identified,
and drop or inject attacks detected accordingly.
\system performs this analysis 
in the collector following \Alg{detection}, as explained next.

The sampled packets arriving at the collector,
a hash of the packet, 
their sampling locations, and arrival times, 
are stored in a list ordered by time. We will refer to this
list as the \emph{history}. Packets are processed after spending 
\emph{max\_delay} seconds in the history, where \emph{max\_delay} 
is a time interval (Round Trip Time) defined to ensure that all other samples of the packet have arrived. 

At the collector, the processing of a packet includes, first, 
constructing the trajectory of the packets, using a policy oracle. 
The policy oracle returns the path (trajectory) that suits the packet. 
Then, considering the switch hash assignments along the path,
all other expected samples of the packet in the history are searched for.
To detect packet modifications,
hashes of the samples are compared to ensure no modifications.
If a sample is 
missing for some switch~$s$, we distinguish between the following two cases
depending on the location of~$s$ along the path and relative to other sampling switches
along the path:
\begin{enumerate}
\item \emph{Path suffix only:} If the packet was sampled by~$s$ but not 
by switches after~$s$, we report a drop attack.
\item \emph{Path prefix only:} If the packet was sampled by~$s$ but not
by switches before~$s$, we report an injection attack.
\end{enumerate}

While the mechanism to identify injection and
dropping events are similar, the severity of, and reaction to these two events
may differ. 
In particular, while injections may occur rather rarely ``by accident'',
benign packet drops do.
Accordingly, for drops arising individually and without
statistical patterns, no alarm should be raised.
To deal with the ephemeral
hash value assignments and avoid false positives, 
we introduce a grace period around dynamic updates.

\begin{algorithm}[t]
	\begin{algorithmic}[1]
		\Require hash assignments~$\Ac$, switches~$S$, incoming event (packet-In) queue~$Q$, network policy oracle~$Policy$
		\State~$History \gets ()$ \Comment\emph{empty sorted list}
		\State~$t_0 = time()$ \Comment\emph{current time}
		\While {$true$}
			\State~$timestamp,pkt,s \gets Q.get()$\Comment\emph{blocking get}
			\State~$History.add(timestamp,pkt,s)$
			\If {$timestamp - t_0 < RTT$}
				\State \emph{continue}
			\EndIf
			\While {$History.min() < timestamp - RTT$}
				\State~$timestamp`,pkt`,s` \gets History.get\_min()$
				\State~$h` \gets hash(pkt`)$
				\State~$path \gets Policy.get\_path\_suffix(pkt`,s`)$
				\For {$s``\in path$}
					\If {$h`\in \Ac(s``)\:\&\&\:(pkt`,s``)\notin History~$}
						\State \emph{Report Drop of~$pkt`$ between~$(s`,s``)$}
						\State \emph{Break}
					\EndIf					
				\EndFor
				\State~$h \gets hash(pkt)$
				\State~$path \gets Policy.get\_path\_prefix(pkt,s)$
				\For {$s``\in path$}
					\If {$h\in \Ac(s``)\:\&\&\:(pkt,s``)\notin History~$}
						\State \emph{Report Injection of~$pkt$ between~$(s`,s``)$}
						\State \emph{Break}
					\EndIf
				\EndFor
			\EndWhile
		\EndWhile
	\end{algorithmic}
\caption{Detection}
\label{alg:detection}
\end{algorithm}

\noindent \textbf{Parallel Trajectory Construction.}
While the sampling approach implemented by~$\system$
is efficient and effective, in order to scale the system further,
we can easily parallelize~$\system$ by executing the
detection algorithm in multiple threads or by
using multiple collectors that execute the detection algorithm.
To increase the throughput of \Alg{detection}
using multiple threads,
the samples received by the collector can be evenly distributed across the
threads.
To increase the throughput using multiple collectors,
the samples from the switches can be sent to different collectors.
Such partitionings are highly scalable: 
sample dependencies are restricted to a single packet.

\section{Analysis of Detection Probability}\label{sec:analysis}
This section presents a formal analysis of 
the detection probability for a single attacker
in~$\system$ under
different attacks, and in different scenarios.
We begin by describing our approach 
for analyzing the probability to detect various
attacks based on a pairwise static assignment
distribution. Following that, we describe the
detection probabilities for a single packet
attack, and flow-based attacks.
Given that the assignment distribution is static,
in the next section, we will
report on prototype experiments which evaluate
and extend our insights to time-varying assignment distributions.

\subsection{General Approach and Notation}
Observe that if either the original or (one of the)
actual route(s) taken by a packet traverses
at least two sampling points, 
$\system$ \emph{guarantees} detection.
This is obvious for simple routing attacks,
i.e., packets which are rerouted, 
injected, dropped, or mirrored. 
However, this also holds for sophisticated
attacks which involve changing packet
contents. For example, consider
a scenario where sampling is based on
TCP checksums and assume 
that an attacker modifies the packet in
such a way that  TCP checksum
remains unchanged. Even in this case,
$\system$ ensures detection:
with the unmodified checksum,
the packet is sampled and sent
entirely to the controller, which
can (bit-by-bit) compare the packets,
and detect modifications.

In addition to the attacked packets, whether the attacker
reports the samples, or not, for the attacked packets affects
the detection probability.
It is better for a single attacker
to report samples for packets that are dropped,
and not report samples for packets that are injected.
In the analysis that follows, we assume that the
attacker follows this strategy with resepect
to reporting samples for attacked packets.

We now focus on the pairwise assignment scheme and
assume a payload packet hashing function. We
consider three types of attacks: 
(1) single packet attacks (drop and inject), (2) dropping an entire flow, 
and (3) injecting a new flow.
We will use the following notations:
\begin{itemize}
	\item $n$: the total number of switches
	\item $p$: the same sampling ratio of the switch
	\item $PS(x,y)$: the number of hash range pairs involving switches from
two sets of sizes~$x$ and~$y$, i.e.,~$PS(x,y)=x\cdot y$
	\item $\widetilde{PS}(x)$: the number of pairs involving at least one switch 
from a set of~$x$ switches, i.e.,~$\widetilde{PS}(x)=x\cdot(n-x) + {x \choose 2}$.
	\item $PS^-(x,y)$:  the number of pairs involving switches of 
a set of size~$x$ but no switches from a second set of size~$y$, i.e., 
$PS^-(x,y)=\widetilde{PS}(x)-PS(x,y)$
	\item $(B,A)$: the number of switches along a path before~($B$)
and after~($A$) the attacking switch respectively, i.e., the attack location
\end{itemize}

\subsection{Single Packet Attack}

In the single packet drop attack, the attacker drops a single
packet from a flow. Similarly, in the single packet injection attack,
the attacker injects a single new packet into a flow.
As we will see, the probability of
detecting a single packet drop attack
and a single packet inject attack, is the same.

To compute the
detection probability, we assume that the attacked packet belongs
to a flow along a single path.
The location of the attack along the
path is expressed by~$(B,A)$ which stands for the number of switches along
the path before and after the attacking switch respectively.
Moreover we assume that all switches have the same sampling ratio,~$p$.
Therefore, each pair of switches that share an assignment contains~$p/(n-1)$
of the hash space $|H|$ (independently at random from the other pairs).

Recall that a detection occurs if the packet is sampled by 
at least one switch before and at least one switch 
after the attack location.
By definition, there are exactly
$PS(A,B)$ pairs
surrounding the attack location~$(B,A)$:
we will refer to these pairs as \emph{explicit pairs}.

However, we observe that the actual detection probability
is slightly higher than what we would obtain
by focusing on the explicit pairs alone. Due to 
a \emph{birthday paradox},
it is probable that other pairs surround
the attack location, as they happen to have the same assignment
before and after the attack location. We will refer to these pairs as 
\emph{implicit pairs}.

\begin{figure}[t]
    \begin{center}
        \includegraphics[width=.9\columnwidth]{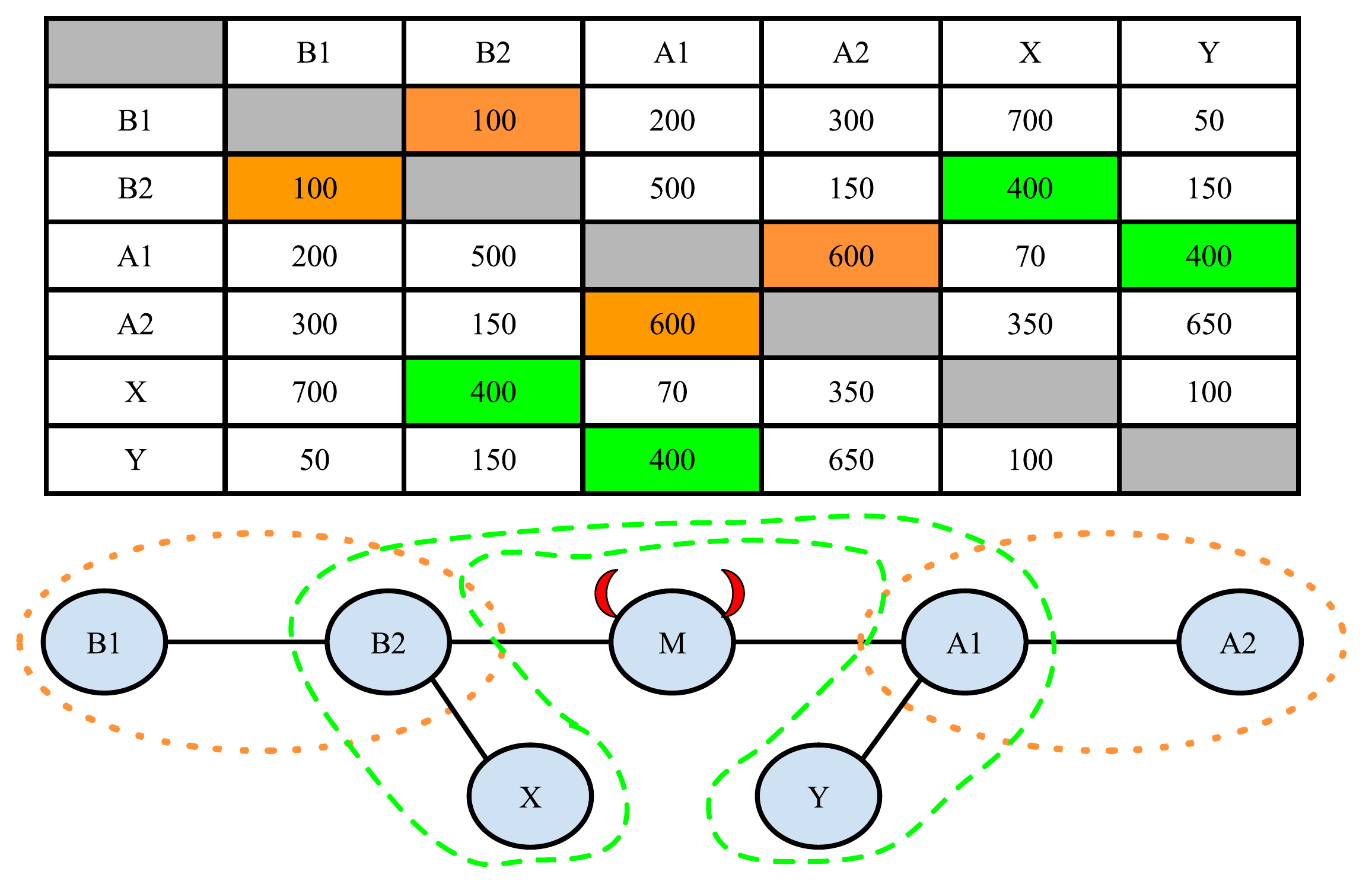}
    \end{center}
    \caption{Illustration of explicit and implicit pairs. 
An example network topology is shown below a table indicating the
hash assignments for each node in the network. The table's
rows and columns indicate the nodes. The assignment for a pair of
nodes is given by the cell that intersects the chosen nodes.
M is the attacker and its position is given by (2,2).
B1 and B2, and A1 and A2, form two explicit pairs, indicated by
the dotted orange lines and the orange cell in
the topology and table resp.
B1 and B2 share the hash value 100, and A1 and A2 share the hash
value 600.
B2 and X, and A1 and Y, also form explicit pairs. However,
the two pairs are assigned the same hash value of 400, making them
an implicit pair. The implicit pair is indicated by the dashed green
line and green cell in the topology and table resp.
To keep the figure simple, not all explicit pairs are shown.}
    \label{fig:explicityImplicitPairs}
\end{figure}

Fig.~\ref{fig:explicityImplicitPairs} visualizes the difference
between explicit and implicit pairs: 
implicit pairs
arise if B2 and X, and A1 and Y 
are assigned the same hash value.

To evaluate the number of implicit pairs, we need 
to compute the collision probability between two sets of 
random variables~\cite{collision}, 
where the (maximal) set sizes are given by~$PS^-(B,A)$ and~$PS^-(A,B)$, 
and the random variables can assume any of~$|\Hc|$ discrete values (or~$|\Hc|/\delta$ considering assignment granularity).
Following work by Yancey~\cite{collision2}, we use~$COL(x,y):=x\cdot y/|\Hc|$ 
as an approximation for the collision of two sets of random assignments 
of sizes~$x$ and~$y$.

The detection probability,
$\pwdp$, then equals the probability that the packet is sampled by one of the 
pairs surrounding the attack location. 
Considering the number of explicit and implicit pairs, we obtain that 
\begin{equation}
\pwdp = 1-{\left(1-\frac{p}{n-1}\right)}^{PS(A,B)+COL\left(PS^-(B,A),PS^-(A,B)\right)}
\end{equation}
We note that if the attacker does not report a sample for
a dropped packet, then the probability of detection is
slightly higher, as the assignments shared between the
attacker, and switches before and after it aid in detection.
Similarly, if the attacker reports a sample for an injected packet,
the chance of detection is equal to that of the drop and the
attacker not reporting the sample.

\subsection{Flow Drop Attack}

Assume now a scenario where an attacker drops the entire flow.
Obviously, also here, the detection probability depends on the
location of the attacker~$(B,A)$ along the flow's path.
Also of importance is the packet hash function used
to detect this attack.
In particular, hashing only the header 
results in the same values for all packets in the flow, 
and a constant detection probability~$\pwdp$ (as for a single packet described
for the single packet attack),
regardless of the number of dropped packets. 
In contrast, hashing the packet's payload leads to more random (per packet)
values, thereby making the detection of each packet an (ideally)
independent experiment, each with success probability ~$\pwdp$.
Therefore, 
the detection probability of the entire attack describes a geometric distribution, 
and the expected number of dropped packets till detection is~$1/\pwdp$.

\subsection{Flow Injection Attack}\label{sec:flowInjectionAnalysis}

We next examine attacks injecting new flows.
It is easy to see that if the injected flow's packets
have uniformly distributed hash values (as assumed for
the original flow),
then the detection probability 
is like that of the drop attack. The expected number of injected
packets till detection is~$1/\pwdp$ in this case.

However, if we consider a more sophisticated
attack wherein the attacker may craft all of
its packets to hash to the same value from
the packet hash function, then the attack
can either be detected with the first packet
from the injected flow, or never. The 
initial (and static)
hash assignment of the switches is a crucial
factor to detect such an attack.

By dynamically configuring the switch pairs
hash assignments
with new random values and at
random times (following a memoryless Poisson distribution),
the detection
probability may be improved.
The update rate should be high
enough to introduce many new
assignments, at the same time it should keep
the number of values updated low:
alerts resulting from the
recently updated assignments
are temporarily suppressed (to avoid
false positives).

The detection of the sophisticated flow inject attack
will occur when the Poisson process of updates results in  
an assignment surrounding the attack position
which includes the hash value used in the injected
flow. We can distinguish between three cases, 
depending on whether the crafted value is assigned before
the attack point, after the attack point, or not at all.    
The derivation of a closed-form expression for the expected detection
time is difficult, 
as there are multiple parameters to consider,
such as the duration of the attack, the update
rate, the update size and the rate of traffic flowing
through the path. 
The duration of the attack
is the most important parameter: at one end of the spectrum
lies the static case, at the other end of the spectrum,
if the assignments change frequently
between injections, the attack is equivalent
to using different hash values and therefore similar
to the flow drop attack.

\subsection{Collusion}

Our analysis generalizes to collusion, as follows. 
We assume that multiple switches collude, e.g., communicate with each 
other (to coordinate a later attack), and do not report each others' samples.
To analyze such behavior, we can adapt our above analysis.
We simply compute the detection probability considering a single
attacker at a time, ignoring the other attackers along the path,
effectively reducing the path length for an attacker under analysis.

\section{Prototype Implementation}\label{sec:prototype}

To demonstrate the feasibility of our approach,
to validate and extend the results of our formal analysis,
and to investigate the potential performance overheads,
we implemented a prototype of \system{}.

We use ONOS-1.4 (implemented in Java) as our controller (collector) and implemented
\system{} as an application on top of it. We used the
OpenFlow 1.3 specification implemented by ONOS for
our prototype. ONOS offers the many
functionalities and services required to implement \system{},
such as multi-threading,
path service, flow rule service, device service,
group service,  and
packet service.

\begin{figure}[t]
    \centering
    \includegraphics[width=0.9\columnwidth]{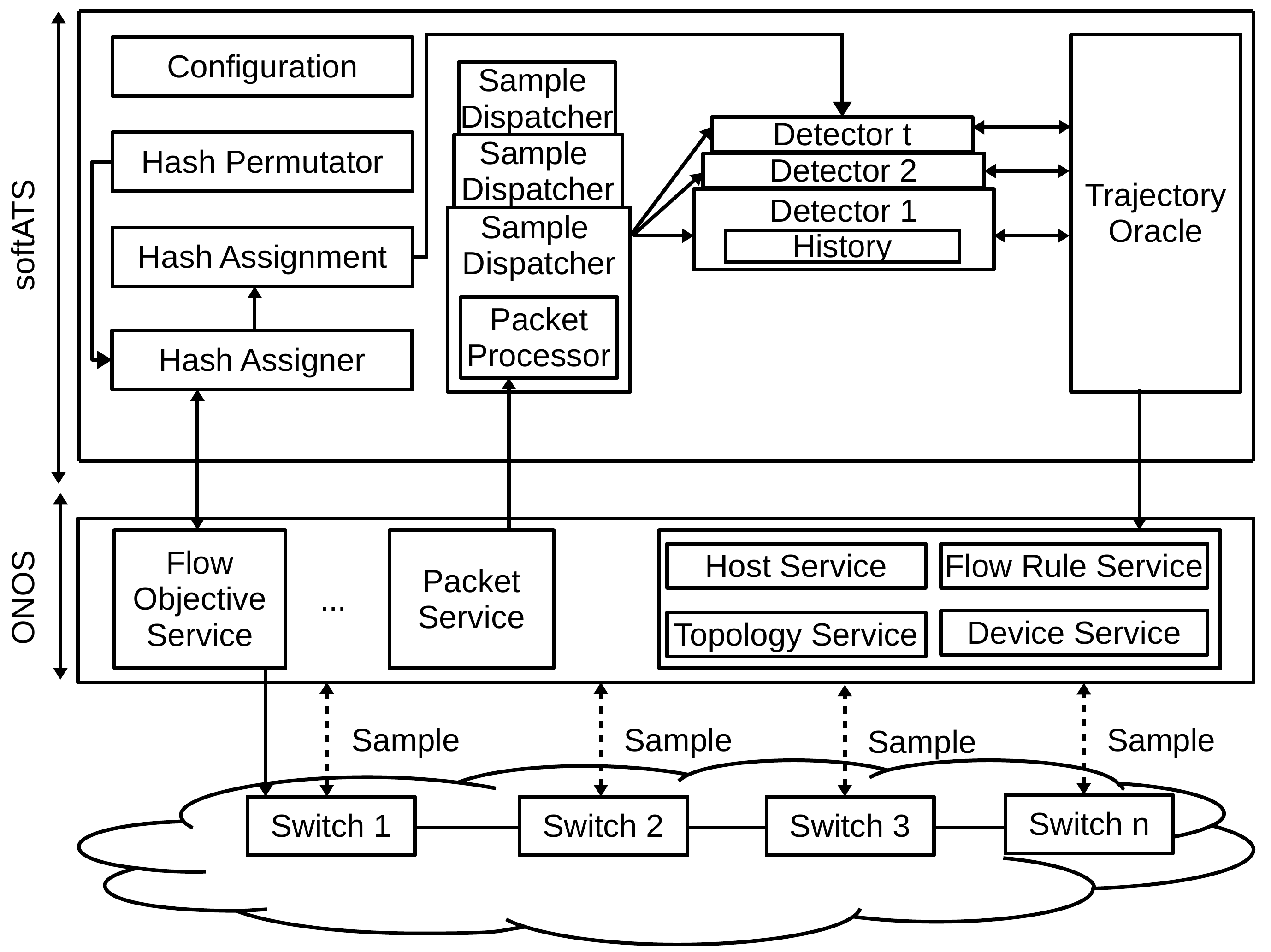}
    \caption{A high-level overview of the \system{} software
architecture.
The \emph{Hash Assigner}, executes
\Alg{assignment}, stores them in \emph{Hash Assignment}, and
uses the Flow Objective Service or Group Service
to install sampling rules on the switches.
A received sample is processed by the \emph{Sample Dispatcher}
and sent to the appropriate \emph{Detector} thread.
The \emph{Detector} uses the \emph{Hash Assignment}, and
path information obtained from the \emph{Trajectory Oracle}
(which has a global network view)
to detect whether an attack has occurred or not,
and report accordingly.}
    \label{fig:softATSdesign}
\end{figure}

Fig.~\ref{fig:softATSdesign} illustrates the architecture
of our implementation.
It is made up of five main components: Configuration,
 Hash Assignment, Sample Dispatcher, Detectors and
a Trajectory Oracle. In the following we will
elaborate on each of the components.
After that, we will briefly describe the data plane
components of \system{}.

\subsection{Configuration}
The different parameters of \system{} such as the sampling
ratio, detector threads, pairwise assignment or
independent assignment, etc. can be configured via
the ONOS Command Line Interface (CLI).

\subsection{Hash Assignment}
The hash assigner executes 
the pairwise assignment of hash values
(\Alg{assignment}) and
utilizes the flow objective service
to assign sampling flow rules to the switches
when payload hashing is used. For
packet header hashing, the group service
is used to assign flow and group table
rules.
The hash assignments are then stored
in a data structure that keeps track
of the hash assignments for all the switches.
The hash assignments are later used
by the detector thread(s) to carry out the
detection algorithm
(\Alg{detection}).
For dynamic assignments,
the hash permutator executes 
the permutation of hash assignment values
(\Alg{permutation})
in its own thread
following a Poisson distribution~\cite{poisson}.
Its average
update rate can also be configured
via the ONOS CLI. It passes the new
assignment to the hash assigner to
have the assignment pushed to the switch(es)
and have the hash assignment data structure
updated.

\subsection{Sample Dispatcher}
The packet service in ONOS receives samples
from the switches as \emph{Packet-Ins}. It strips
away the OpenFlow header and passes the
Ethernet frame (packet) to the
sample dispatcher's packet processor (FIFO) queue of \system{}.
The sample dispatcher
always takes the first sample from the
queue and
dispatches the sample to the appropriate
detector's history queue.
The logic for the dispatcher and the detector
threads are based on how one wants
to distribute the sampling load.
In our case, we divide the hash space $|H|$
evenly across the detector threads.
The dispatcher simply checks which
detector thread is responsible for
the particular sample (according to its hash)
and places the
sample into the detector's history queue.

\subsection{Detectors}
Each detector thread
executes the detection algorithm (\Alg{detection}) and therefore
has its
own history queue which allows it to operate
independent
of the other threads. There are
no dependencies on the other threads.
The only concurrency requirement
on the history queue is that
operation on it be synchronized.

\subsection{Trajectory Oracle}
The network policy oracle implemented
as the trajectory oracle in our prototype,
returns the path suffix and path prefix to
the detector by obtaining host, device and path
information from ONOS services.
If an attack is detected by the detector,
an alert is generated in the log messages
of ONOS.

\subsection{Data Plane}
For the data plane, we use Open vSwitch (OvS)
as it readily implements OpenFlow 1.3 and supports
packet header hashing via \emph{group tables}.
By default, OvS, ONOS and OpenFlow do not support
matching the checksum field of TCP/UDP packets
in a flow rule~\cite{shahbaz2016pisces}.
Therefore, we use the currently supported VLAN tag field
to match packets for sampling.
We tag all our traffic and populate
it with the value from the checksum field
truncated to the tag field's size.
Since the tag field
is only 12 bits and the checksum field is 16 bits,
we effectively reduce $|H|$ from 65535 to 4096,
which we deem acceptable for our proof-of-concept
prototype and experiments.

\section{Experimental Evaluation}\label{sec:eval}

Using our prototype and realistic traffic workloads,
we have conducted extensive experiments on a Clos topology (a
``fat-tree'').
We first report on our experimental setup.
Following that,
we report on the observed detection times,
and then discuss the 
throughput of computing trajectories in \system{}.
Finally, we study the resource overheads introduced by $\system$.

\subsection{Setup}
\label{sec:setup}
\begin{figure}[t]
    \centering
    \includegraphics[width=0.9\columnwidth]{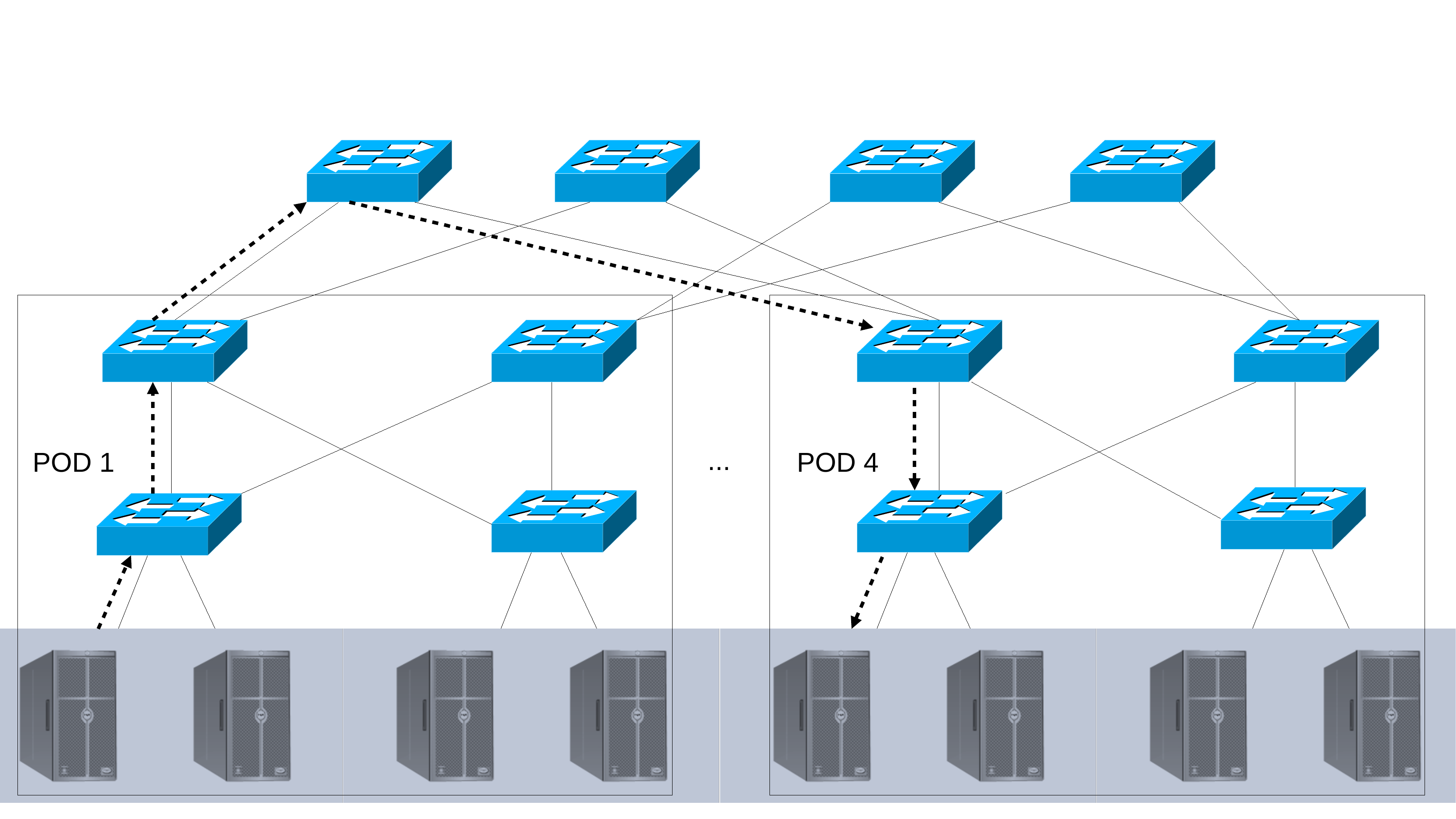}
    \caption{Topology used for detection time, trajectory throughput and
sampling overhead experiments of \system{}.
The traffic between hosts flows as per the arrowed and dotted line i.e., traffic
flows from a host in Pod 1 to a host in Pod 4.}
    \label{fig:topo}
\end{figure}

We consider a Clos/fat-tree topology with $k=4$, using the Ripcord
platform for Mininet~\cite{ripl,Casado:EECS-2010-93}.
To generate realistic network traffic,
we leverage the Lawrence Livermore Berkeley
National Laboratories (LLBNL) traffic~\cite{pang2005first} traces.
Using a custom script,
we extract only the internal subnet-to-subnet flows 
from the traces,
then extract the TCP checksums from those flows
and replay traffic using those checksums within our
topology using tcpreplay at 100 pps as shown in Fig.~\ref{fig:topo}.
Considering the concepts defined in Sec.~\ref{sec:softatssys},
our default parameters for \system{} are as follows:
\textbf{Detectors~$t$:} 1;
\textbf{Packet Hashing Function~$h$:} payload dependent;
\textbf{Sampling Ratio~$p_s$:} 0.4\%;
\textbf{Assignment Granularity~$\delta$:} 1;
\textbf{Redundancy:} Pairwise;
\textbf{Average Hash Update Rate:} 2s and
\textbf{Hash Update Size:} 2.
The number of hash assignments per switch
and sampling ratio are chosen
such that every switch in the network forms one pair
with every other switch. 
Recall from Sec.~\ref{sec:softatssys},
~$p_s = \frac{|\Ac(s)|}{|\Hc|}$.
Since our topology
has 20 switches, we have $|\Ac(s)|=20$,
and $|H|=4096$, we get $p_s=$$\sim$$0.4\%$
as the sampling ratio.
In the following, we will explicitly state any changes
to the default parameters and traffic.

\subsection{Detection Time}

To validate and complement our formal analysis of the mean detection time, 
we conducted a set of experiments studying the detection time 
using real network traffic. 
In particular, we evaluate our detection algorithm
(Algorithm~\ref{alg:detection}) in the presence of 
a single attacker, and two colluding attackers.
We focus on the
flow drop, and flow injection
attacks for the single attacker.
For the colluding attackers,
we focus on the flow injection attack.
In both attacks (flow drop, and flow inject),
we use a static and time-varying
sampling distribution for our detection algorithm.

\noindent \textbf{Setup:} 
Since the sampling ratio impacts the detection time,
in addition to the the default parameters specified in Sec.~\ref{sec:setup}, we use the following:
\textbf{Sampling Ratio~$p_s$:} 0.9\% and 1.3\%.

\subsubsection{Single Attacker}
\label{sec:singleAttacker}
We first describe our methodology, results and analysis
for a single attacker.

\noindent \textbf{Methodology:} We evaluate the effectiveness of
detecting the flow drop and flow injection attacks when
the attacker is the aggregate switch and when the attacker
is the core switch along a single path
in our topology.
In the flow drop, the switch drops all packets
from a flow along a path. Furthermore, it reports the samples,
if any, for dropped packets.
In the flow inject,
the switch injects a new flow along the path
of an existing flow. Additionally, it does not report samples
for the injected packets.
Both attacks are easily
emulated via OpenFlow flow rules on OvS.
We count the number of packets that
are sent in a flow till an alarm is raised by \system{}
and then stop. We perform 100 such trials for
\textbf{each attack} and \textbf{each attacker}.
We use the packet count as a metric instead of
time to remain independent of the traffic rate:
what are ``realistic rates'' depends on the context
(e.g., data center vs ISP).

\begin{figure}[t]
    \centering
    \includegraphics[width=0.99\columnwidth]{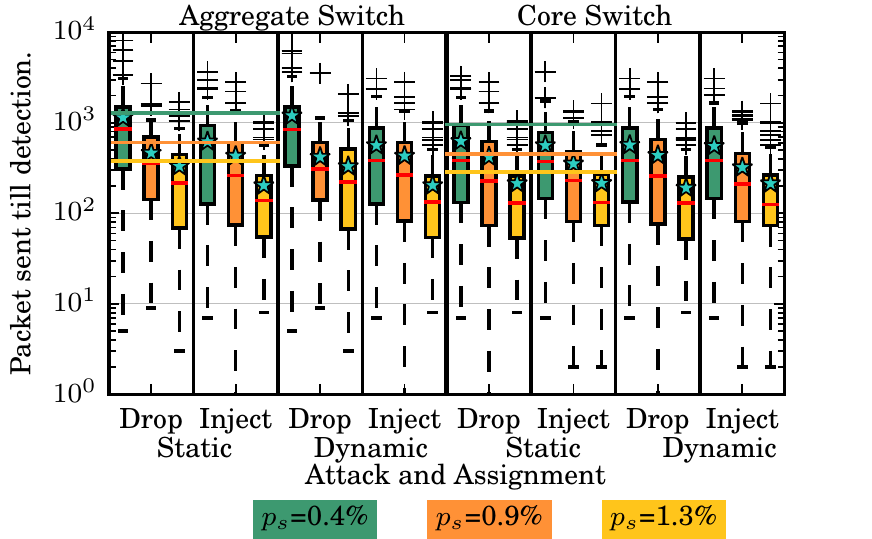}
    \caption{Average no. of packets to detect the attacks at different positions using
the pairwise assignment in the static and dynamic setting with an average
update rate of 2s and an update size of 2 and varying sampling rates.}
    \label{fig:experimentalDetectionTimeResults}
\end{figure}

\noindent \textbf{Results and Analysis:} Fig.~\ref{fig:experimentalDetectionTimeResults}
shows the results from the experiments.
The theoretical means are represented as solid colored
lines. The box plots represent the empirical
results: the star represents the mean,
and the solid red line represents the median.
First, the figure confirms our theoretical results:
the empirical results are very close to the 
analytical mean for the detection time.
However, while in theory the
flow inject attack
takes as long as the flow drop attack, empirically
it is detected sooner. 
This is possible for two reasons. First our analysis of the detection
probability is conservative, i.e., we do not consider
the pairs between the attacker, and other benign
switches along the path, that can in fact
help in detection.
Second, the hash assignments are not purely
independent, i.e., for the chosen sampling ratios,
each switch had exactly one, two and three pairs
shared with every other switch in the network.
This increased the number of unique pairs
surrounding the attacker, thereby increasing
the detection probabilities.
Furthermore, we observe a lot of variance in detecting
the various attacks.
We suspect this is due to the non-uniform distribution
of the TCP checksum field in the traffic used.
Partridge et al.~\cite{Partridge:1995:PCC:217391.217413}
measured the performance of
the TCP checksum, and identified that for small sized
packets, the distribution is skewed for the UNIX
file system. Nonetheless, based on our experiments,
the results work in favor of \system{}.
The sampling ratio~$p$ also improves the detection,
roughly linearly: by doubling~$p$ we detect
the attack in half the expected number of packets.
The position
of the attacker also influences the detection, i.e.,
it takes fewer packets to detect the malicious
core switch than the malicious aggregate switch.
This is due to the fact that there are more explicit and implicit
pairs surrounding the
core switch than the aggregate switch, hence improving
the detection probability.
In Sec.~\ref{sec:extension}, we analyze the relationship
between the detection probability, the path length,
and the attacker's position in the path.

\subsubsection{Colluding Attackers}
\label{sec:colludingAttackers}
Now we describe our methodology, results and analysis
for two colluding attackers.
\noindent \textbf{Methodology:} We evaluate the effectiveness of
detecting the flow injection attack (analogous to mirroring) when
two aggregate switches collude:
the two switches collude to not report samples for
all packets injected.
We emulate that by modifying the specific sampling flow
rules on the two switches to 
not report matching packets to the controller.
In this attack, the benign traffic flows from
Pod 1 to Pod 4 (as shown in Fig.~\ref{fig:topo}.
However, the injected traffic flows from Pod 1 to
Pod 3.
The remainder of the methodology is the same as
the single attacker in Sec.~\ref{sec:singleAttacker}.

\begin{figure}[t]
    \centering
    \includegraphics[width=0.99\columnwidth]{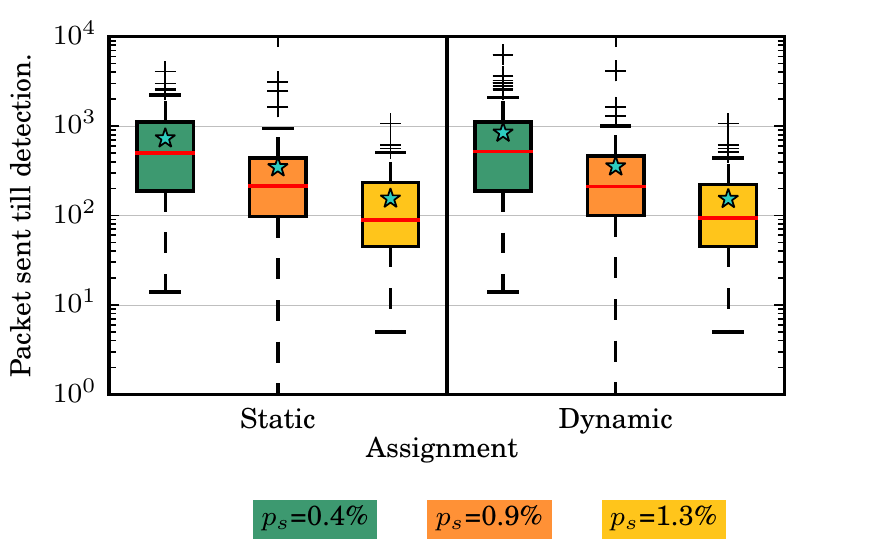}
    \caption{Average no. of packets to detect a flow inject attack when two switches collude, using
the pairwise assignment in the static and dynamic setting with an average
update rate of 2s and an update size of 2 and varying sampling rates.}
    \label{fig:experimentalDetectionTimeResultsCollusion}
\end{figure}

\noindent \textbf{Results and Analysis:} Fig.~\ref{fig:experimentalDetectionTimeResultsCollusion}
shows the results from the collusion experiment.
The figure shows that for two colluding switches
not reporting samples for all injected packets,
\system{} requires less than 1000 packets to
detect the attack.
Naturally, we observe that with fewer benign switches,
it takes more packets to detect an inject attack.
Using the dynamic assignment, we observe that the values
chosen do not result in significant changes to the detection
compared to the static assignment.

\subsection{Detection Throughput}
\label{sec:trajTput}

Next, we study the number of trajectories per second which can be 
analyzed in parallel, i.e., the detection rate. Recall that \system{} is multi-threaded
(Sec.~\ref{sec:softatssys}).

\noindent \textbf{Setup:} The evaluation was carried out on a
64 bit Intel Core i7-3517U CPU @ 1.90 GHz with
4GB of RAM. 
We are interested in measuring the throughput of computing
trajectories, therefore, we use multiple detector threads.
Hence,
in addition to the default parameters as mentioned in Sec.~\ref{sec:setup}, we use the following:
\textbf{Detectors~$t$:} 2, 4, 6 and 8.
Note that each detector thread is an independent thread within \system{},
that can compute trajectories and detect attacks.

\noindent \textbf{Methodology:} To measure the detection throughput (trajectories/s) of
the multi-threaded \system{},
we record the total time taken for a single sample
to be dispatched to its respective detector and
for the detector to complete the trajectory
computation.
Each detector thread computes 1k trajectories, from
which we
compute the mean throughput for \emph{t} detector threads,
which is given by Eq.~\ref{eq2}. We then repeat the measurements for
different CPU core counts (1, 2 and 4/hyper-Threading).
\begin{equation}
throughput=\frac{\textit{t}}{\textit{mean(DispatchTime + DetectionTime)}}
\label{eq2}
\end{equation}

\begin{figure}[t]
    \centering
    \includegraphics[width=0.8\columnwidth]{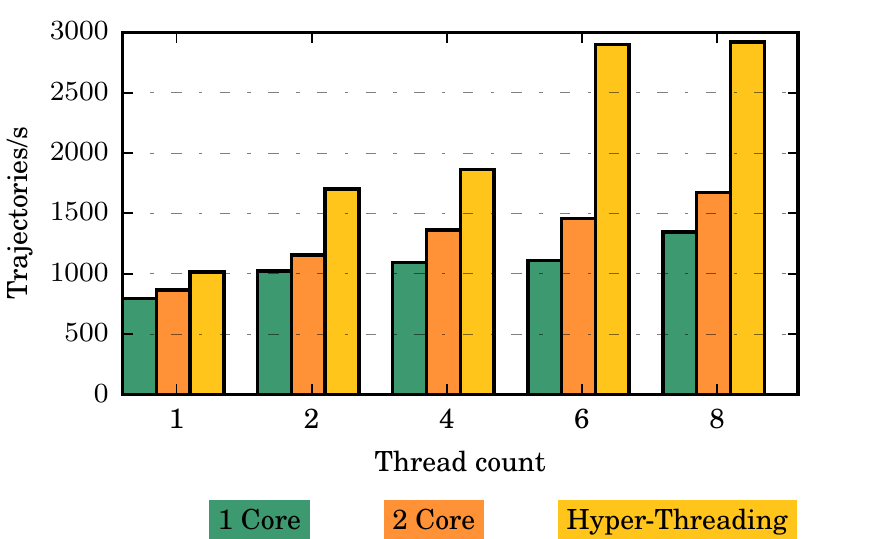}
    \caption{Comparison of SoftATS mean throughput (trajectories/s) with varying
thread counts and CPU cores.}
    \label{fig:trajectoryThroughput}
\end{figure}

\noindent \textbf{Results and Analysis:} The results obtained are shown in
Fig.~\ref{fig:trajectoryThroughput}. 
It is clearly evident that there is an increase in the
throughput with an increase in detector threads.
However,
the throughput does not increase
linearly with the thread count and the number of cores.
Nonetheless, there is a performance improvement.
Using 8 threads and a hyper-threaded CPU, the throughput
is close to 3000 trajectories/s which is $\sim$3 times that
of using a single thread on a single core CPU.
The non-linear increase can be attributed to
the History list
(see Detection~\Alg{detection}), where read and
write operations are synchronized.
Furthermore, we suspect that:
(i) ONOS itself requires CPU cycles
to manage and maintain itself and the switches,
(ii) I/O interrupts cause the
operating system to preempt \system{} and ONOS, 
and (iii) the CPU architecture
used for this experiment uses hyper-threading
(Intel's implementation of simultaneous multi-threading),
influencing the results non-deterministically.
Nevertheless, the results lend credence to the use
of multiple detection threads to achieve
high detection rates and high availability.
Furthermore, the use of multiple threads on multiple
collectors can substantially increase the
detection throughput.

\subsection{Overhead of Sampling}
\label{sec:overhead}

Although \system{} has benefits in adversarial
environments, there are potential overheads.
In the following we measure and analyze
the overhead of sampling at the
(ONOS) controller/collector, and the switch
(OvS). We first characterize the overhead in terms of
CPU and memory usage at the controller.
Next, we measure the overhead, if any, sampling introduces
to the forwarding throughput and latency of different sized
network packets.

\subsubsection{Overhead at the Controller}
\label{sec:overheadController}
For overhead at the controller, we are
primarily interested in measuring the
CPU and memory utilization of \system{}.
We begin by describing our experimental setup,
followed by our methodology, and then conclude
with the results and analysis.

\noindent \textbf{Setup:}
We use the same system from
the detection throughput experiment for characterizing
the overhead of sampling at the controller.
For \system{}, we use the default parameters as mentioned
in Sec.~\ref{sec:setup}.

\noindent \textbf{Methodology:} To characterize the CPU and memory
consumption at the collector, we first measure the resource
consumption without \system{}, then
enable a static configuration of \system{}, and finally
we enable the dynamic configuration of \system{}.
In all the cases, we replay bi-directional traffic at 100 pps
between 6 pairs of hosts, i.e., each pod receives 3 flows,
for 10 minutes. 
We measure the CPU and memory consumption using \emph{top}
every second with the controller pinned to a single core.

\begin{figure}[t]
    \centering
    \includegraphics[width=0.99\columnwidth]{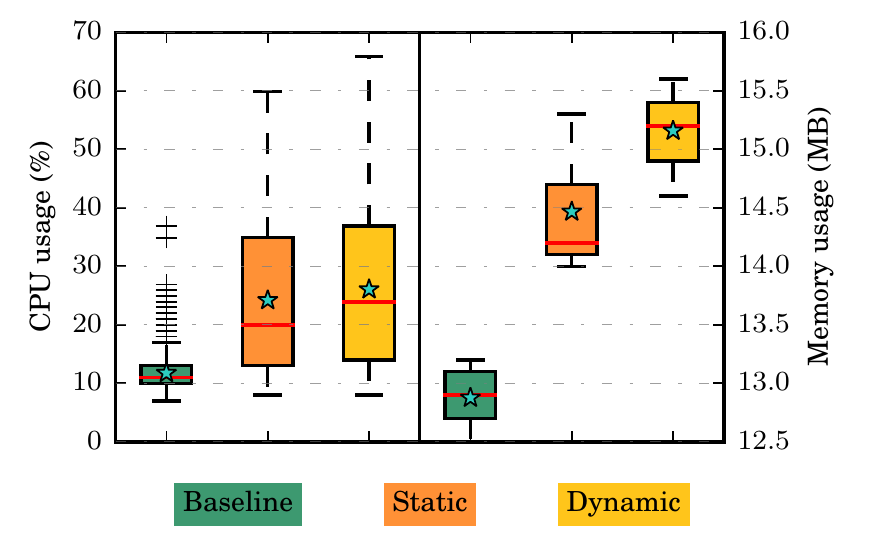}
    \caption{Resource consumption of \system{}.}
    \label{fig:overhead}
\end{figure}

\noindent \textbf{Results and Analysis:} Fig.~\ref{fig:overhead} shows
the results.  Naturally, \system{} does increase CPU and
memory consumption on average compared to the baseline usage.
The CPU increase is due to \system{} dispatching a sample to the
detector thread(s) and computing the trajectory.
The memory increase by approximately 1.5 MB is due to the
data structures used by \system{} to store and process samples and
hash assignments for switches.
50\% of the CPU usage for the static and dynamic cases is within
12-35\% usage, and  50\% of the memory usage is
within 14-15.5 MB: both are within
acceptable boundaries for scenarios where \system{} executes on
a controller that manages the network. 
Furthermore,
the dynamic configuration, even with an
average update rate of 2s, does not increase the mean
CPU and memory usage by much, compared to the static
configuration. This implies that the benefits
of the dynamic configuration can be reaped with
minor overheads in terms of CPU and memory.

If \system{}
were to be used as a stand-alone system or with
multiple collectors, these overheads are 
very low. In fact, as we have demonstrated previously
in Sec.~\ref{sec:trajTput}, the detection throughput
can be increased by using several detector threads, and
the load can be distributed
across several collectors, thereby reducing the
resource consumption on individual collectors.
Hence, we see the overhead at the controller to be a
small trade-off for the benefits of a
highly parallel, distributed and performant detection
system such as \system{}.

\subsubsection{Overhead at the Switch}
\label{sec:overheadSwitch}
Having investigated the overhead at the controller,
we now measure the overhead of sampling at
the switch. The two main metrics we are interested in
here are: throughput, and latency.
Using a traffic generator and hardware timestamps, we are able to accurately
and precisely measure the impact of sampling on
the throughput and latency of \system{} at the switch.
In the following, we describe our experimental setup for evaluating
the impact of sampling on the forwarding performance.
Due to our limited resources, we evaluate the
overhead using a software switch, namely, Open vSwitch.

\begin{figure}[t]
    \begin{center}
        \includegraphics[width=0.89\columnwidth]{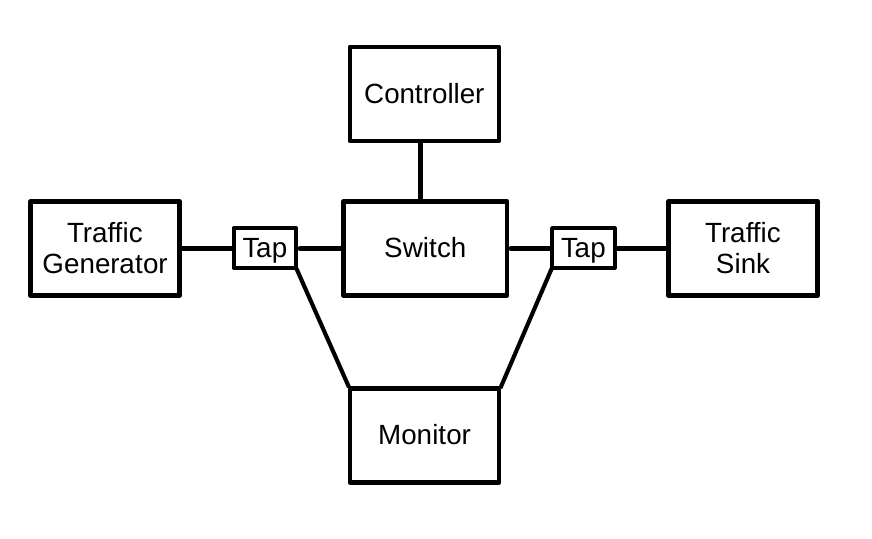}
    \end{center}
    \caption{Topology used to measure the forwarding performance
of OvS.}
    \label{fig:forwardingTopology}
\end{figure}

\noindent \textbf{Setup:}
Since we are interested in the forwarding performance of
the switch when \system{} is used,
we use one server running OvS. Therefore,
the topology in this section uses only one switch
and one controller.
In addition to the default parameters from Sec.~\ref{sec:setup},
we use the following:
\textbf{Sampling Ratio~$p_s$:} 0\%, 0.9\% and 1.3\%.
We only use the static assignment scheme in this experiment.
To measure the throughput and latency of OvS,
we use five systems as shown in Fig.~\ref{fig:forwardingTopology}.
We use a Traffic Generator---to
generate specific sized packets, and replay them at
specific rates to OvS.
OvS forwards packets to the Traffic Sink.
OvS is also connected to the Controller
via a dedicated interface.
Using passive network taps
on the ingress and egress links of OvS,
we collect the traffic at the Monitor system.
Except for the Monitor, all other systems use two
2.5 GHz Dual Core AMD Opteron CPUs with
16 GB of RAM, and four Intel Gigabit Network
Interface Cards (NICs).
The Monitor uses two dual-core 3.7 GHz Intel Xeon CPUs
with hyper-threading enabled and 16 GB of RAM.
We use an Endace DAG 10X4-P card to capture network
packets on the Monitor. Each interface of the
card has a receive queue configured with 1 GB.
This provides us with accurate, and highly precise uni-directional
measurements, independent of the host system's (Monitor) resource
utilization.
All the systems run Ubuntu 14.04.
For OvS, we use ovs-2.3.2 compiled using the default
configure script with gcc.
OvS also has a kernel-based
fast-path, we use version:~20D84E92C1F09E01E1586EE
running on a Linux kernel version 4.6.5.

Furthermore, we configure OvS with one rule
for sampling, and one rule for forwarding, apart from
the two default flow rules installed by ONOS's proxyarp,
and reactive forwarding application.
The sampling rule matches VLAN tags only,
while the forwarding rule matches packets based
on the Ethernet source and destination MAC addresses.
The transmitted packets, are UDP packets encapsulated
in IP packets. Ethernet is used to encapsulate the
IP packets.

\noindent \textbf{Methodology:}
Our objective is to measure the uni-directional
forwarding performance overhead, if any, of OvS
using our sampling scheme.
Therefore, we first measure the forwarding
performance of OvS without our scheme, i.e.,
\textbf{Sampling Ratio~$p_s$:} 0, and then
with sampling, i.e.,
\textbf{$p_s$:} 0.4\%, 0.9\% and 1.3\%.

To measure the throughput,
we transmit a constant stream of UDP packets at a specific
rate and measure the rate at which the stream is
forwarded by OvS at the Monitor.
To unearth performance bottlenecks quickly, we
use packets per second (pps) as the metric for
throughput~\cite{jacobson1988congestion}.
To identify the impact of packet sizes on
the throughput, we use three packet (frame) sizes (Bytes)
based on the IEEE 802.3 Ethernet standard for
basic frames~\cite{ethernet8023}:
64, 512 and 1518.
For each packet size, we transmit packets for
330 seconds at a constant rate. We then
reset the setup and change the transmit rate.
The transmit rate starts at 
10 kpps and goes up to 100 kpps, in steps of 10 k.

To measure the latency,
we transmit specific sized packets and measure
the time taken for OvS to forward them.
We do so for packet sizes (Bytes) based on the Ethernet
standard: 64, 128, 256, 512, 1024, 1280 and 1518.
For each trial, we send 10500 specific sized packet
at 100 pps. We discard the
first 500 packets to eliminate any system artefacts.

\begin{figure}[t]
    \centering
    \includegraphics[width=0.99\columnwidth]{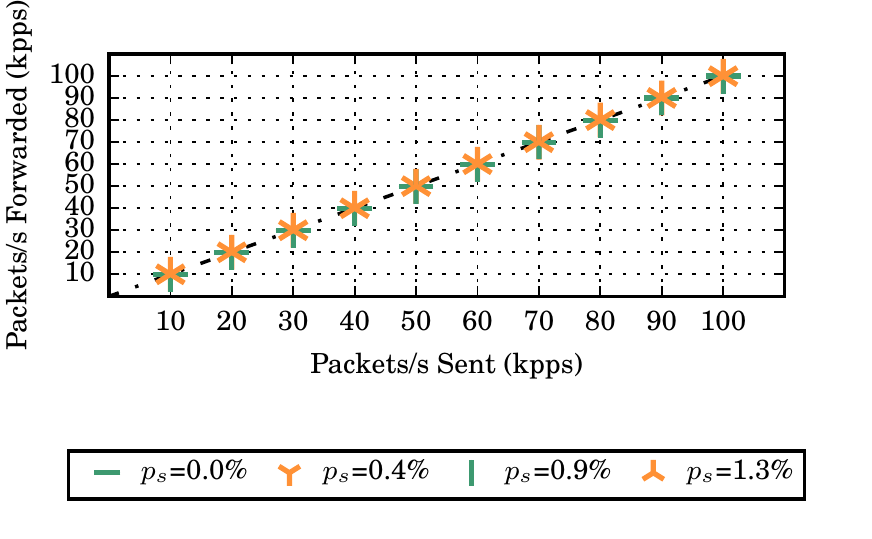}
    \caption{Forwarding throughput of 64 byte packets for different sampling ratios.}
    \label{fig:tputOverhead64}
\end{figure}

\begin{figure}[t]
    \centering
    \includegraphics[width=0.99\columnwidth]{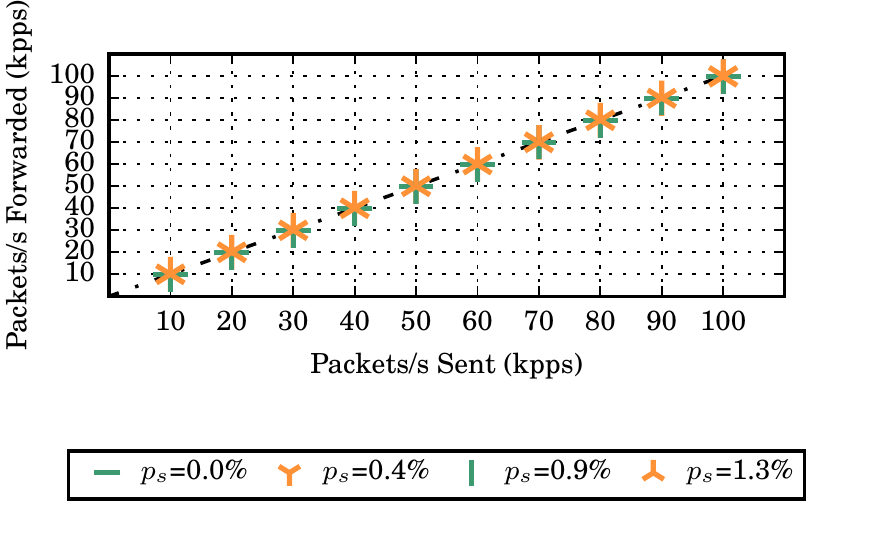}
    \caption{Forwarding throughput of 512 byte packets for different sampling ratios.}
    \label{fig:tputOverhead512}
\end{figure}

\begin{figure}[t]
    \centering
    \includegraphics[width=0.99\columnwidth]{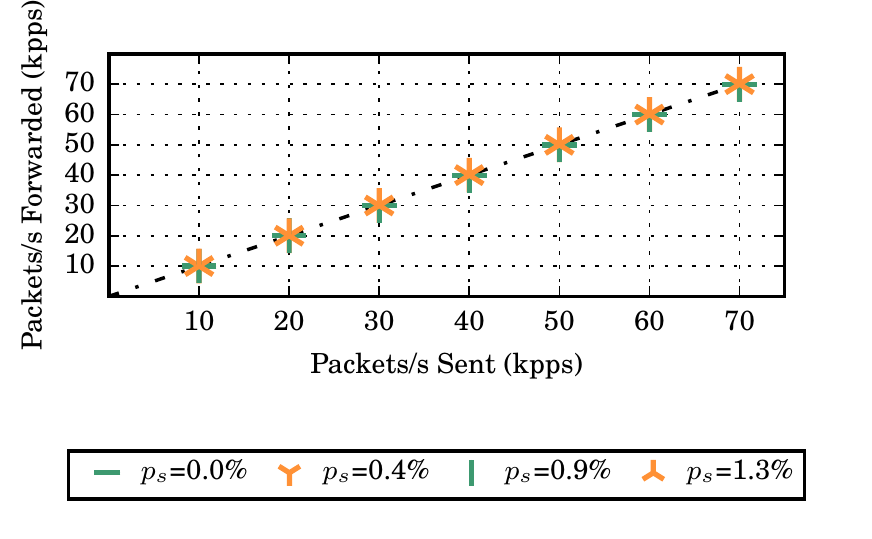}
    \caption{Forwarding throughput of 1518 byte packets for different sampling ratios.}
    \label{fig:tputOverhead1518}
\end{figure}

\noindent \textbf{Results and Analysis:} Fig.~\ref{fig:tputOverhead64},
~\ref{fig:tputOverhead512}, and~\ref{fig:tputOverhead1518} show
the overhead of sampling for different sized packets.
For 64 and 512 byte packets, OvS could reach 100 kpps transmission
rates, however, for 1518 byte packets, the peak transmission rate
was only 70 kpps. The results show that for the sampling
ratios chosen, the forwarding throughput of OvS is not impacted.

\begin{figure}[t]
    \centering
    \includegraphics[width=0.99\columnwidth]{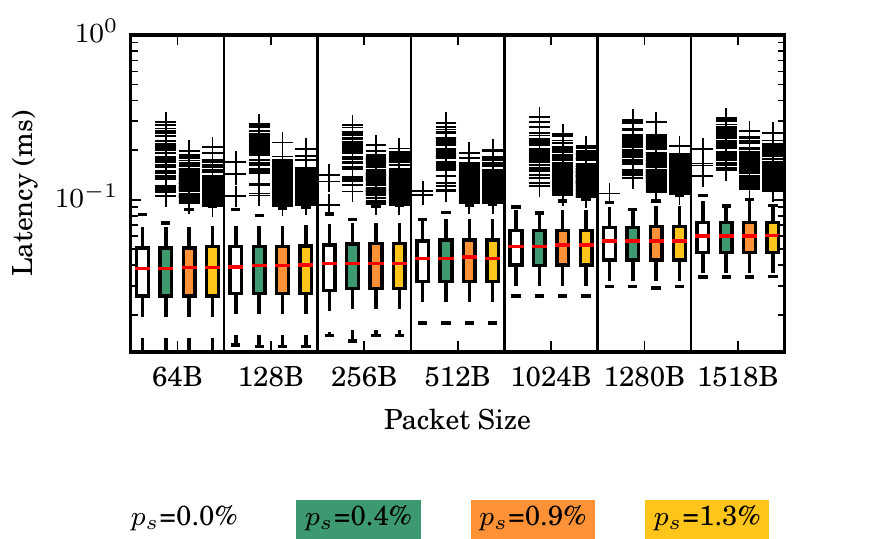}
    \caption{Forwarding latency of different sampling ratios using various sized packets at 100 pps.}
    \label{fig:latencyOverhead100pps}
\end{figure}

Fig.~\ref{fig:latencyOverhead100pps}
shows the forwarding latency of OvS for
different sized packets without and with sampling.
It is evident that sampling introduces many outliers.
This is because sampling increases the
packet processing pipeline compared to plain
forwarding. When a packet matches
the sampling rule in OvS it has to go through
OvS's slow path. Therefore, it has to go
from the kernel-space to user-space for processing.
The action of sending the packet to
the controller increases the packet processing.
Lastly, the packet is matched against two tables
until it is forwarded,
as opposed to only one table in the baseline.
From the figures we can also see that
regardless of sampling,
for 1024 byte packets and higher, the latency increase
by roughly 25\%.

\section{Discussion}\label{sec:discussion}

Our approach raises opportunities for further
improvements, but also comes with limitations. 
We will discuss some of
them in more detail in the following.

\subsection{Sampling Extension to the Edge}\label{sec:extension}

\begin{figure}[t]
	\includegraphics[width=.45\textwidth]{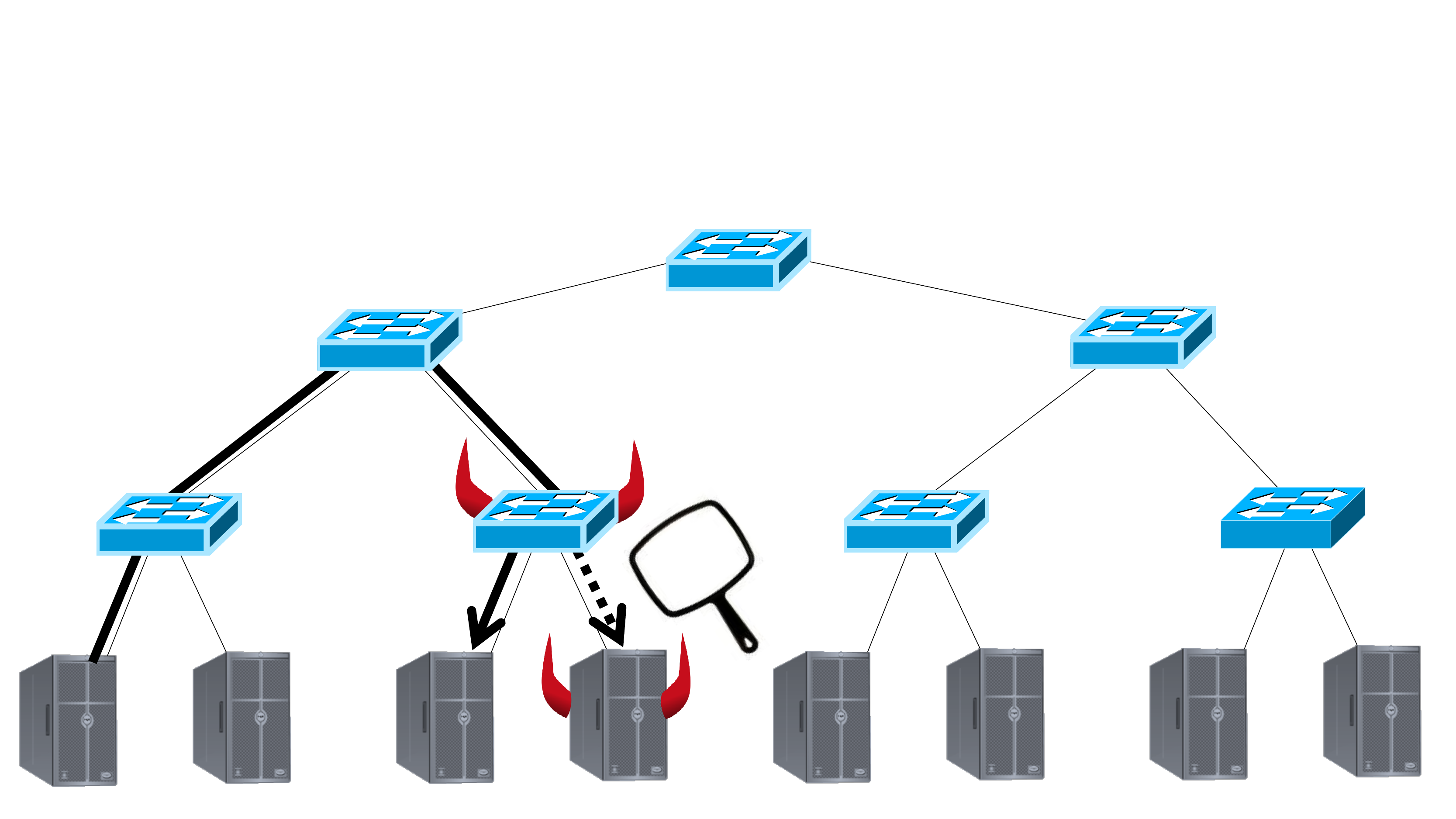}
	\caption{Detecting attacks by the top-of-the-rack switch
	is hard.}
	\label{fig:tor-attacks}
\end{figure}

Our adversarial trajectory sampling scheme comes with the fundamental 
limitation that the network edge must be trusted:
if a packet does not even enter the sampling network, it is impossible
to detect a routing error related to it.
For example, consider the scenario depicted in 
 Figure~\ref{fig:tor-attacks}:
As there are no sample points on the route after the
malicious ToR switch, it is impossible to detect
the mirrored packet to the malicious host.
 
 \begin{figure}[t]
	\includegraphics[width=.99\columnwidth]{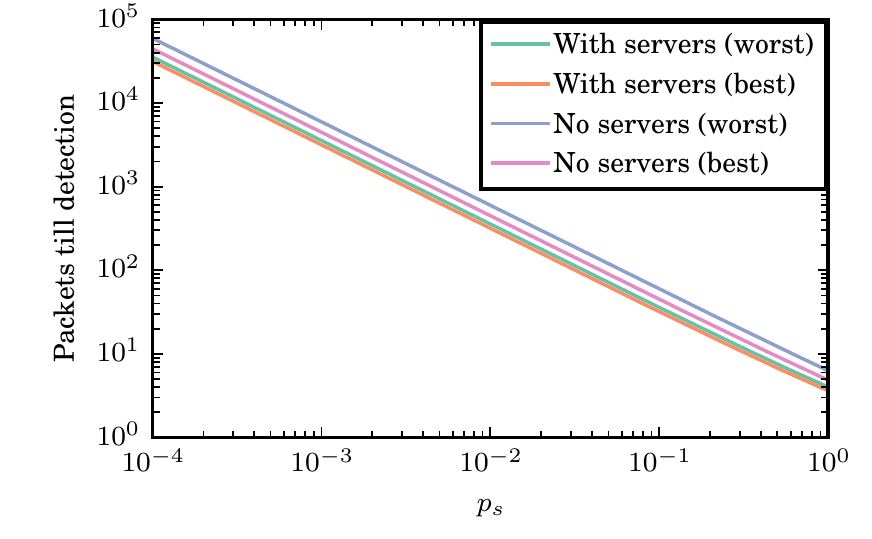}
	\caption{A comparison of the number of packets to detection with and without including the edge servers in \system{} in the best and worst case.}
	\label{fig:compareEdgeDetection}
\end{figure}

However, the problem can be alleviated if the sampling scheme is extended to the edge.
For example, 
in cloud management systems such
as OpenStack, Open vSwitch (OvS) is
used on the virtualized servers
which is already OpenFlow enabled.
Assuming that the
virtualized server's host OS
can be trusted,
OvS
(or the networking stack
on the host OS)
can also participate in our
scheme to detect misbehaving
ToR switches.

Fig.~\ref{fig:compareEdgeDetection} compares the packets till
detection when the servers are used, and not used in \system{}.
We indicate detecting the aggregate switch as the worst case,
and detecting the core switch as the best case. This is
because the aggregate switch has the least number of pairs
surrounding it, and the core switch has the most number of
pairs surrounding it.
Clearly, there is a benefit in using the servers with respect
to detecting attacks. The servers improve the detection time
by a factor of roughly 2x. By adding the servers to \system{},
the total number of switches increase from 20 to 36 which
increases the explicit pairs across the network.
Furthermore, including the servers increases the number of
switches surrounding the attacker by two, thereby increasing
the detection probability.

One interesting challenge of this 
approach regards how to securely connect OvS
with the collector:
this connection needs to be implemented in-band, 
via other (possibly) malicious (ToR) switches.
 
An alternative approach to 
extend the security
to the edge of the network could be to use a robust combiner approach~\cite{rob-comb,disn16netco}, i.e., by
 leveraging physical redundancy (e.g., switches of different vendors or switches
 which were fabricated in different countries).

\subsection{Applying SoftATS to Other Networks}
So far we have used a data center network as our
primary topology of \system{}. We
observe that \system{} is by no means limited
to such networks.
It can also be applied to
wide area networks (WAN) and
service provider networks. Admittedly, to use
\system{} in the Internet, all nodes across
end-points need to participate in our scheme.
Accomplishing this, is no small feat and is beyond
the scope of this paper. However, it is
still feasible for \system{} to be used
within an administrative domain such as
an internet service provider for example.

We expect no modifications to \system{} when
used in a service provider network with the
exceptions of obtaining centralized control,
and global routing information. In fact,
we observe that using \system{}, one can
obtain fine-grained monitoring over the network,
as the sampling can be applied at switches (L2)
and routers (L3) in unison.
In addition, we expect WANs and service provider networks
to have long paths.
Therefore, in the following we analyse
the relationship between the detection probabilities
and the path length using ATT's network
topology.

\begin{figure}[t]
	\includegraphics[width=.99\columnwidth]{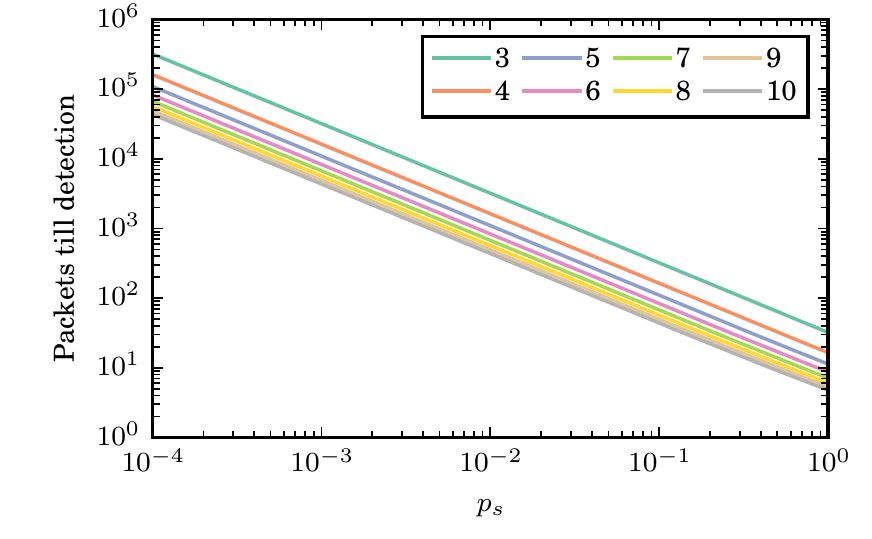}
	\caption{A comparison of the number of packets to detection for the worst case in ATT's network for various path lengths.}
	\label{fig:worstCaseDetection}
\end{figure}

\begin{figure}[t]
	\includegraphics[width=.99\columnwidth]{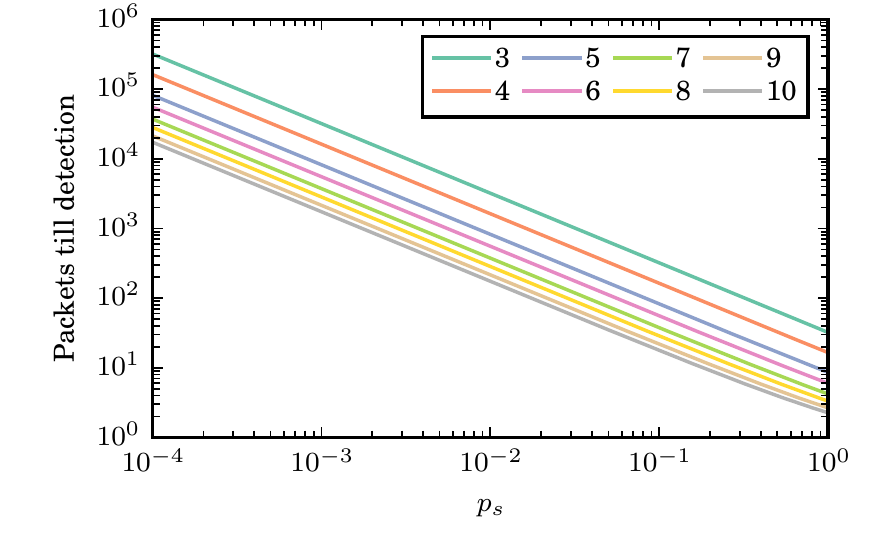}
	\caption{A comparison of the number of packets to detection for the best case in ATT's network for various path lengths.}
	\label{fig:bestCaseDetection}
\end{figure}

The ATT network we use~\cite{attTopo}, comprises of
a total of 54 nodes with a median path length
of 7~\cite{sats}. Such a network gives us room
to explore how the detection
probability and path lengths interact. This
was not possible in the Clos topology we used
for our evaluation.
Fig.~\ref{fig:worstCaseDetection} and~\ref{fig:bestCaseDetection},
show the relationship of the detection probability
to path lengths for the worst and best case resp.
The worst case for a given path is the attacker's position
that has the fewest pairs surrounding it, i.e.,
the node adjacent to the edge node.
The best case is the attacker's position that
has the most pairs surrounding it, i.e.,
the middle node on the path.
From both figures it is clear that 
the packets till detection and the sampling ratio
grows linearly.
Furthermore, it appears that increasing the
path length reduces the packets till detection
exponentially only till a certain length beyond
which there is little benefit in long
paths. Nonetheless, it shows that longer paths
and more nodes, do indeed increase the
detection probability.

\subsection{Localization Is Difficult}

While our sampling scheme is useful to
\emph{detect} misbehavior, it is hard to \emph{localize} a misbehaving switch.
To illustrate this, let us consider some examples.
When two switches that are far apart (e.g., the source and destination
switches on a long path), share an assignment
that generates an alert, it is difficult to isolate
the malicious switch:~it could be any of the switches
after the source and before the destination switch.
Localization is particularly hard for
injection attacks: it is difficult to 
determine the misbehaving switch, if the packet is not 
sampled at all the switches.

Nevertheless, each detected misbehavior
gives some clues where things went wrong.
Accordingly, due to our time-varying sampling strategy,
there is always a chance that leveraging tomographic techniques
to perform localization is actually
possible for a concrete event.
Over time, such patterns may become statistically
significant.

\subsection{Attacks on the Collector}

\system{} is built upon the assumption that
it is hosted on a secure system. Therefore, if an attacker
compromises the system that hosts \system{} then
our scheme can be broken. Hence,
appropriate measures (which are beyond
the scope of this paper) must be taken to ensure
that the system hosting \system{} is
sufficiently secured.

The availability of our scheme relies on
the system(s) running the collector(s), the link
capacities between the switches and the collector(s), etc.
Indeed an attacker can overwhelm the collector with
samples and then carry out the attack. However,
there already exist measures to deal with such
problems, e.g., using redundant and multiple
collectors.
By rate-limiting the samples arriving
from the switch, the collector may also be protected
from a dos attack.
To ensure availability of the scheme
and the network, it is best for \system{}
to run on dedicated and multiple controllers.

The attacker can reorder transit packet thereby
influence the order of the samples reported to
the collector. If the delay in receiving the
sample is beyond the $max\_delay$, then \system{}
will throw two alerts. One drop attack alert for the first sample
and one inject attack alert for the second sample.
It will not miss detecting such an attack.

\section{Related Work}\label{sec:relwork}

Being able to measure routes and traffic 
is critical for controlling and engineering
a communication network~\cite{jrex}. Accordingly,
the topic has been studied intensively in the
past. 
One generally distinguishes between indirect and direct measurement
methods. 
An indirect measurement method relies
on a network model and network status information to infer
the traffic route (e.g., IGP and BGP routing state).
  Indirect measurement methods suffer from
the uncertainty associated with the physical and logical state of
the network. 
In contrast, direct methods rely on
direct observations of traffic at multiple points in the network.
Trajectory sampling~\cite{ts,bck-ts} is an example of the latter.

Many operating systems support
the \emph{traceroute} and \emph{ping}~\cite{traceroute1,traceroutes}
command to display the route (path) and 
to measure transit delays of packets across an Internet Protocol (IP) network.
These tools are based 
on active test packet injection
and are attractive for their parsimonious use of
data plane resources.
However, it is well-known that 
ping, traceroute, and counter-based solutions~\cite{whisper}
fail in adversarial settings~\cite{jrex}:
For example, a malicious switch 
may fake responses to provide the
impression that probing packets traverse the network properly.
Both active~\cite{traceroute1,traceroutes} and passive probing~\cite{ts,bck-ts} 
are vulnerable to forged reports. In general, any
adversarial behavior that can maliciously misreport
samples, or delete or modify information contained
in packet headers, essentially breaks a scheme based on packet
labeling or tagging~\cite{pathquery,pathtracer}.

Trajectory Sampling~\cite{ts,bck-ts} is
a direct  measurement
method based on a consistent sampling approach
allowing to reconstruct the path, at the collector,
given a pseudorandom subset of packets through
the corresponding network domain.
The approach is attractive as most switch vendors today implement some form
of
packet sampling (e.g., Netflow~\cite{netflow}).
However, while trajectory sampling 
may still perform well in the context of faulty networks~\cite{unrel-ts}
and networks with different flow sizes~\cite{csamp},
it is insufficient in non-cooperative and adversarial
environments.

Interestingly, while over the last years,
much research was conducted on how to secure
routing protocols on the control plane~\cite{secroute1,secroute2,secroute3,sbgp,Subramanian:2005:DSM:1195621},
providing authenticity and correctness of topology propagation and route computation, 
the important question of how to secure the data plane
has received much less attention so far.
In fact, until very recently, 
researchers did not know whether it is possible
to build a secure path verification
mechanism~\cite{icing}. 
Many existing systems like 
VeriFlow~\cite{veriflow}, Anteater~\cite{anteater} and
Header Space Analysis~\cite{header-space} 
rely
either on flow rules installed at switches or on data plane
configuration information to perform their analysis. 
This information can easily be manipulated in malicious
settings.
Providing fundamental 
properties like path consent and
path compliance~\cite{icing} are
usually based on cryptographic
techniques: expensive operations
in high speed networks.
Even more challenging than inferring the
routes along which 
certain packets actually travelled is 
to test where packets \emph{did not travel}~\cite{alibi-routing}.

Goldberg et al.~\cite{jrex} propose an interesting approach for
end-to-end path quality monitoring;
while this approach is suitable also in very general settings,
it comes at the cost of introducing extra information
in packets and requiring stronger hash functions.
In an SDN setting where communication between switches
and controllers are out-of-band and secured,
opportunities for simpler solutions such as ours
are introduced.

Our approach is particularly motivated by the introduction of
software-defined networks. 
While SDNs are known to introduce many flexibilities, 
also in terms of security, they also introduce new
threats. For instance, Yu et al.~\cite{Yu:2014:DCT:2620728.2620739}
presented a distributed traffic monitoring scheme for SDNs,
and
FleXam~\cite{FleXam} is a sampling extension for monitoring 
and security applications in OpenFlow.
NetSight~\cite{netsight} leverages SDN
to trace entire packet histories (without sampling),
by collecting them ``out-of-band'',
and
CherryPick~\cite{cherrypick} uses 
packets to carry information of SDN paths ``in-band''
(namely, a 
subset of links along the packet trajectory);
however, these protocols struggle with drops and are not robust to
malicious switches. In particular,
the information CherryPick adds to the header along the path
is only verified at the end of the path.
Bates et al.~\cite{bates-2014-forensics}
use SDN networks (plus some middleboxes)
to observe the data plane behavior, even in the presence
of malicious switches.
The traffic engineering flexibilities of SDN have also been exploited 
to perform secret sharing~\cite{shlomi}.
Zeng et al.~\cite{traceroutes} use SDN to
test the forwarding and policies in the network
by generating and actively probing the data plane
across the network.

Desai et al.~\cite{desai2012packet} propose
a hash-based delay sampling technique to detect
switches misforwarding packets. Their threat model
and objectives are closely related to ours. However,
their sampling scheme and detection algorithm
are different from ours. They require three switches
to sample the same set of packets and their detection
algorithm depends on the state of the switches
buffers for a chosen path. We can incorporate
their method of sampling i.e., choose triplets
instead of pairs, but our detection is
based on trajectories and not switch states.

Yu et al.~~\cite{yu2013software} describe OpenSketch,
a generic and efficient sketch-based measurement framework for SDN
data planes. They developed
APIs and sketches to make generic SDN measurements
alleviate the control plane programming complexity
operators face.
Additionally, they present a prototype using
NetFPGA to demonstrate the feasibility, applicability and
overhead of their approach.
Since their framework is designed to be parallel to the
packet processing pipeline, there is no performance
impact to the forwarding. However,
matching OpenFlow flow rules and
sending samples 
to the controller 
are not feasible.
Nonetheless, their framework
can be modified to implement our scheme.

Bu et al.~\cite{7524333} introduce accurate and
efficient algorithms to detect and troubleshoot
flow rule and flow table faults in the data plane.
They do so by using probe packets through the data plane.
This approach is vastly different from 
ours. We rely on a passive approach, therefore we do not
send probe packets through the data plane.
Moreover, our objective is to detect
incorrect packet trajectories rather than
incorrect forwarding rules and tables that may cause
incorrect trajectories.
Furthermore, our threat model does not restrict
the cause for misforwarding to only flow rules and flow
priorities even though it can be used to do so.

The paper most closely related to ours is by
Lee et al.~\cite{sats}. The authors
present a smart generalization of trajectory sampling
where hash values are shared between a subset
of switches, rendering it hard for an adversary to avoid
detection. The authors also present a first simulation
of the possibility to detect packet drop, modifications
and substitution attacks.
Our paper builds upon this work in multiple respects.
We initiate the study of injection attacks and observe that
if combined with drop attacks, injection attacks introduce a number
of more sophisticated attacks. Accordingly, we extend
and formally and empirically analyze detection algorithm
guarantees under various misbehaviors, 
including 
injections, mirroring, rerouting, or modifications of headers
and/or payloads.
Unlike prior work, our 
hash assignment algorithm is completely random, 
eliminating bias for pairs and their assigned hash values.
Moreover, our detection algorithm
does not rely on an aggregation of trajectories and
counters. Instead, we use the collector's (controller's) global view
of the network (i.e., the network policy oracle) to compute every sampled packet's
trajectory which gives us a per sample detection
accuracy rather than an aggregate.
We demonstrate that our detection algorithm 
can be parallelized using multiple threads (and possibly
multiple collectors).

In general, our work is motivated  by the insight that SDNs
provide an ideal environment to implement
our algorithms, and we present an OpenFlow prototype accordingly.
We understand~$\system$ as a flexible framework which 
supports a wide range of algorithms, tailored towards their
specific settings and threat models.

\section{Conclusion}\label{sec:conclusion}

Today's computer network routing protocols incorrectly assume
that 
switches are non-malicious, and accordingly, 
a sufficiently large portion of the
computer network infrastructure
is vulnerable
to security attacks from compromised switches,
 e.g., containing hardware or software backdoors~\cite{Subramanian:2005:DSM:1195621}.

We in this paper presented a simple and light-weight
adversarial sampling approach based on the software-defined
networking paradigm, which can detect a wide range
of adversarial behaviors, and in different settings,
e.g., in the datacenter but also in the wide-area network~\cite{sdx}. 
Our scheme cannot only detect packet
drops, but also packet injections, rerouting,
or even packet headers and content alterations.

Furthermore, we implemented our \system{} in
the standard OpenFlow protocol, evaluated the
prototype in terms of detection time, and
detection throughput. We also measured
the overhead introduced by \system{} and identified
that at the controller there is a modest increase
in CPU and memory utilization. At the switch, there
is little to no forwarding overhead, making
\system{} feasible in the real world.

We believe that our work opens several
interesting directions for future work.
In particular, our framework allows to experiment
with many alternative algorithms, which can be 
optimized and tailored towards specific
use cases (e.g., enterprise networks) and attacks.
Moreover, we have so far focused on detection only,
and many interesting questions for future research
arise when aiming at the \emph{localization} 
and  \emph{prevention} of
malicious nodes and/or behaviors. 

\noindent \textbf{Acknowledgments:} 
Research supported in part by the German Federal Office for Information Security,
the Danish Villum foundation project \emph{ReNet}, and
the Helmholtz Research School on Security Technologies 
at the German Aerospace Center and the Technical University Berlin.
We would like to 
thank Jens Sieberg 
and Nir Gazit for many 
inputs and feedback on this work. 

\bibliographystyle{IEEEtran}
\balance
\bibliography{refs}
\balance

\end{document}